\documentclass[aps,pra,showpacs,twocolumn,amsmath,amssymb,nofootinbib]{revtex4}
\usepackage{graphicx}
\usepackage{epsfig,color}
\usepackage{bm}
\usepackage{amssymb}
\usepackage{amsmath}
\usepackage{latexsym}
\DeclareMathOperator{\sgn}{sgn}
\newcommand{\vc}[1]{\mbox{\boldmath $#1$}} 
\newcommand{\ind}[1]{_{#1}}    
\newcommand{\indrm}[1]{_{\mathrm {#1}}}    
\newcommand{\timex}{\mathcal{T}_{\ind{\Lambda}}}
\newcommand{\timesix}{\mathcal{T}_{\ind{6m}}}

\newcommand{\timesixr}{\mathcal{T}_{\ind{6r}}}
\newcommand{\snphi}{\sin\varphi}   
\newcommand{\csphi}{\cos\varphi}   
\newcommand{\esi}{\vc{e}_{\ind{\sigma}}}
\newcommand{\epi}{\vc{e}_{\ind{\pi}}}
\newcommand{\epp}{\vc{e}_{\ind{+}}}
\newcommand{\epm}{\vc{e}_{\ind{-}}}
\begin{document}
\title{Broadly Tunable Compact Noncoplanar X-Ray Cavity for Cavity-Based Free-Electron Lasers}

\author{Yuri Shvyd'ko} \affiliation{Advanced Photon Source,
  Argonne National Laboratory, Argonne, Illinois 60439, USA}
%
\begin{abstract}
  Cavity-based x-ray free-electron lasers (CBXFELs) require tunable
  x-ray cavities to serve a broader range of scientific applications.
  We consider and analyze a compact, tunable, noncoplanar x-ray cavity
  based on six Bragg-reflecting crystals arranged as two three-crystal
  backscattering units in either a mirror-symmetric or a twofold
  rotationally symmetric configuration. Its three-dimensional geometry
  provides a much larger tuning range than planar bowtie cavities
  while keeping a compact transverse footprint. We derive analytical
  expressions for the cavity geometry, crystal kinematics, and
  polarization-dependent Bragg reflectivity in both the intrinsic and
  eigenpolarization bases. For a given Bragg reflection, the ideal
  photon-energy range spans $E_{\mathrm H}$ at backscattering to
  $2E_{\mathrm H}$ at $\theta=\pi/6$, corresponding to 100\%
  tunability. Geometrical and polarization constraints reduce the
  directly usable range to approximately 58\%. Operating in the cavity
  eigenpolarization basis can extend it further, limited only by the
  narrowing angular acceptance of the Bragg reflections. By combining
  fundamental and harmonic diamond reflections, as well as reflections
  from different crystallographic families, the accessible
  photon-energy range can be extended from about 3~keV to
  approximately 18~keV. These results establish noncoplanar
  multi-crystal cavities as a potentially viable route to compact,
  broadly tunable x-ray resonators for future CBXFELs.
\end{abstract}


\maketitle

\section{Introduction} Cavity-based x-ray free-electron lasers
(CBXFELs), including low-gain x-ray free-electron laser oscillators
(XFELOs)~\cite{KSR08,KS09,LSKF11,DDD12} and high-gain x-ray
regenerative amplifier free-electron lasers
(XRAFELs)~\cite{HR06,FSS19,MHH20}, have been proposed as a promising
route toward fully coherent, stable x-ray sources with exceptionally
high average spectral brightness. A CBXFEL comprises a low-emittance,
high-repetition-rate electron beam; an undulator system in which
x-rays are generated and amplified; and an x-ray optical cavity. The
cavity stores and recirculates x-ray pulses, enabling repeated FEL
interactions with fresh electron bunches until saturation is
reached.

Such cavities require low-loss, wavefront-preserving x-ray
optical components, including near-100\%-reflectivity diamond crystals
operating in Bragg reflection~\cite{SSC10,SSB11}, outcoupling elements
such as thin diamond membranes~\cite{KVT16,LPM24} or x-ray
gratings~\cite{MRH23,LPM24}, and aberration-free focusing
lenses~\cite{KSG18} or mirrors \cite{RDL23}.

Lasing of a CBXFEL was recently demonstrated at the European XFEL at
6.952~keV~\cite{RBF26}. A 9.831-keV CBXFEL test system is under
commissioning at LCLS through a joint ANL--SLAC--RIKEN
collaboration~\cite{MAA19,LPM24,LPS24}. Recent developments in this
rapidly evolving field are reviewed in Ref.~\cite{LRHD26}.

The x-ray cavities in these pilot projects employ  fixed-energy
designs optimized for practical implementation. Tunable x-ray cavities
are needed to broaden the practical utility of CBXFELs and to provide
access to a wider range of scientific applications.

Planar tunable x-ray cavities were proposed almost 60 years
ago~\cite{Cotterill68,Cotterill70}. They consist of two sets of $n$
identical crystal reflectors, with the Bragg angle $\theta$ varying in
the range $\pi/(2n)<\theta<\pi/n$. The four-crystal bowtie geometry,
corresponding to $n=2$, was later revisited and studied for
application to CBXFELs~\cite{KS09}. Its ultimate tunability is
determined by the accessible Bragg-angle range
$\pi/4<\theta<\pi/2$. In practice, however, this range is strongly
constrained by the cavity geometry. At approximately
$\theta\simeq 11\pi/32$, the transverse size of the bowtie cavity
becomes comparable to its longitudinal size. Cavities with smaller
Bragg angles are therefore impractical for CBXFEL applications, in
which undulator lengths are typically tens of meters or longer and the
cavity must fit within the available accelerator-tunnel geometry. As a
result, the practical tuning range of the bowtie cavity is limited to
about 11\%.

Here we consider a tunable-cavity design that is transversely more
compact and provides a substantially larger tuning range. This is
achieved by increasing the number of Bragg reflectors from four to six
and arranging them in a noncoplanar, three-dimensional geometry. The
configuration can be viewed as two three-crystal backscattering units,
analogous to a pair of optical retroreflectors. The basic concept and
schematic of such a cavity were introduced more than a decade
ago~\cite{Shv13}, but without a detailed analysis of its optical
properties or tuning range. A related cavity configuration has also
been used in numerical simulations of an
XRAFEL~\cite{FSS19,vdSF21}.

\begin{figure*}[t!]
\setlength{\unitlength}{\textwidth}
\includegraphics[width=0.99\textwidth]{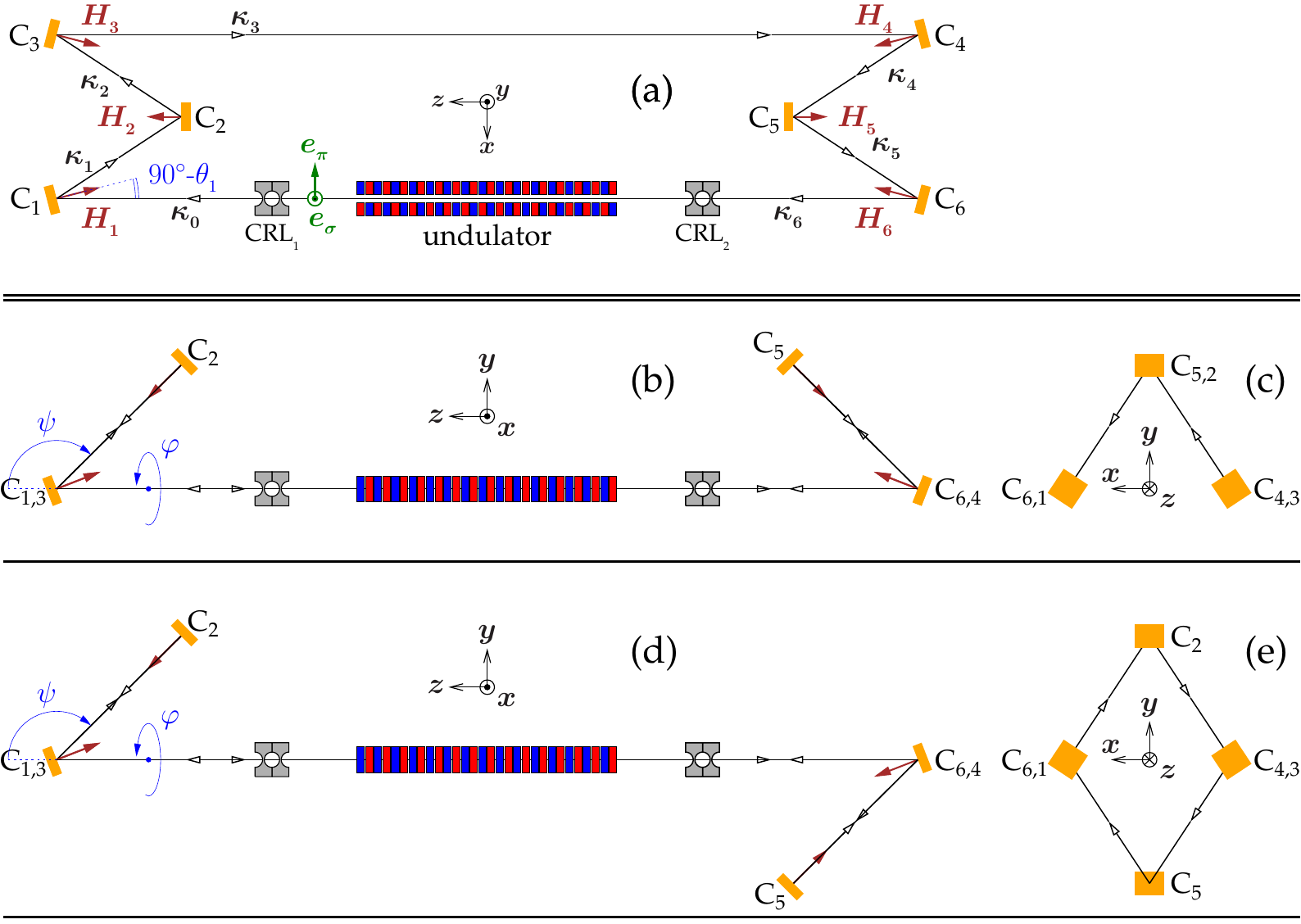}
\caption{Optical layouts of a tunable compact six-crystal noncoplanar
  x-ray cavity in two symmetric configurations. (a) Top view. (b,c)
  Front and side views of the mirror-symmetric configuration. (d,e)
  Front and side views of the twofold rotationally symmetric
  configuration. The crystals C$_{\ind{n}}$ ($n=1,2,\ldots,6$) define
  a closed orbit for the circulating x-rays and enable broad
  photon-energy tunability through variation of the Bragg angles
  $\theta_{\ind{n}}$ and the rotation angles $\varphi$ and $\psi$. The
  compound refractive lenses (CRLs) stabilize the x-ray orbit and
  refocus the x-rays onto the electron beam in the undulator. The
  vectors normal to the crystal plates indicate the corresponding
  diffraction vectors $\vc H_{\ind{n}}$. The vectors $\esi$ and $\epi$
  denote the intrinsic polarization basis.}
\label{fig1}
\end{figure*} 

In this work, we present a detailed theoretical study of this
six-crystal noncoplanar cavity. The analysis includes analytical
expressions for the cavity geometry and the kinematics of the
reflecting crystals as functions of Bragg angle and photon energy. We
also derive polarization-dependent Bragg-reflection matrices for the
three-crystal backscattering units and for the complete six-crystal
cavity, both in the geometrically defined intrinsic polarization basis
and in the cavity eigenpolarization basis.

The ultimate photon-energy tuning range of the cavity is 100\%, from
$E=E_{\ind{H}}$ at the backscattering condition $\theta=\pi/2$ to
$E=2E_{\ind{H}}$ at $\theta=\pi/6$. In practice, however, this range
is reduced by several constraints. The first is geometrical: the
longitudinal dimensions of the cavity become too large as
$\theta\to\pi/2$. The second is polarization-related:
cross-polarization mixing becomes large and the common
total-reflection region, as defined within the dynamical theory of
Bragg diffraction \cite{Zach,Pinsker,Authier,Shvydko-SB}, becomes
narrow as $\theta$ approaches $\pi/4$. Together, these constraints
reduce the practical tuning range to approximately 58\% for a given
Bragg reflection.

The cross-polarization-mixing constraint can be mitigated by operating
the cavity in one of its eigenpolarization states, which can increase
the usable tuning range limited only by the narrowing angular
acceptance of the Bragg reflection.  The accessible photon-energy
range can be expanded further by employing both fundamental and
harmonic diamond reflections.

These results establish noncoplanar multi-crystal cavities as a
potentially viable path toward transversely compact, tunable x-ray
resonators for CBXFEL applications in the spectral range from about
3~keV to approximately 18~keV.

\section{Cavity Configurations and Kinematics}

CBXFEL optical cavities comprise flat Bragg-reflecting crystal mirrors,
which define a closed orbit for the circulating x-rays, and x-ray
focusing elements, such as refractive lenses or mirrors, which
stabilize the orbit and refocus the x-rays onto the electron beam in
the undulator
\cite{KSR08,KS09,Shv13,FSS19,MAA19,MHH20,MRH23,RDL23,LPM24,LPS24,RBF26}. In
this paper, we focus only on the crystal optics. Discussions of stable
and self-consistent cavity modes determined by focusing optics can be
found in Refs.~\cite{LSKF11,QS22}.

\subsection{Cavity configurations}

Figure~\ref{fig1} shows the optical layouts of a six-crystal
noncoplanar x-ray cavity with crystals C$_{\ind{n}}$
($n=1,2,\ldots,6$). The cavity consists of two backscattering units,
C$_{\ind{1}}$--C$_{\ind{3}}$ and C$_{\ind{4}}$--C$_{\ind{6}}$,
arranged either in a mirror-symmetric configuration with mirror plane
$(x,y)$, Figs.~\ref{fig1}(a)--(c), or in a twofold rotationally
symmetric configuration with rotation axis $x$, Figs.~\ref{fig1}(a),
(d), and (e). The cavity enables broad photon-energy tunability
through variation of the crystal Bragg angles $\theta_{\ind{n}}$ and
the crystal rotation angles $\varphi$ and $\psi$. The compound
refractive lenses (CRLs) serve as focusing elements.

In each backscattering unit, the x-ray beam undergoes three successive
Bragg reflections from three individual crystals with diffraction
vectors $\vc{H}_{\ind{n}}$. In the chosen reference frame, the central
wave vector of the incident x-ray beam is
$\vc{\kappa}_{\ind{0}}=\kappa(0,0,1)$, and the diffraction vectors
$\vc H_{\ind{n}}$ in the first backscattering unit are
\begin{equation}
  \begin{aligned}
\vc{H}_{\ind{1}}& = H_{\ind{1}} \left(\, -\cos\theta_{\ind{1}}\csphi ,\, \cos\theta_{\ind{1}}\snphi ,\,  -\sin\theta_{\ind{1}}\,  \right) \\
\vc{H}_{\ind{2}}& = H_{\ind{2}} \left(\,  0,\, -\sin\psi ,\, -\cos\psi\,  \right)\\ 
\vc{H}_{\ind{3}}& = H_{\ind{1}} \left(\, \cos\theta_{\ind{1}}\csphi ,\,\cos\theta_{\ind{1}}\snphi ,\,  -\sin\theta_{\ind{1}} \, \right) 
  \end{aligned}
\label{eq0100}  
\end{equation}
We assume that the first and third reflections have diffraction
vectors of equal magnitude, $H_{\ind{1}}=H_{\ind{3}}$.  The rotation
angles $\varphi$ and $\psi$ are illustrated in Fig.~\ref{fig1}. Here
$\varphi$ is defined as positive for counterclockwise rotation about
the $\hat{\vc z}$ axis.

The  diffraction vectors $\vc H_{\ind{n}}$ ($n=4,5,6$) of the crystals
forming the second backscattering unit  are given by
\begin{equation}
  \begin{aligned}
\vc{H}_{\ind{4}}& = H_{\ind{1}} \left(\, \cos\theta_{\ind{1}}\csphi
                  ,\, \pm \cos\theta_{\ind{1}}\snphi ,\,  \sin\theta_{\ind{1}}\,  \right) \\
\vc{H}_{\ind{5}}& = H_{\ind{2}} \left(\,  0,\, \mp \sin\psi ,\, \cos\psi\,  \right)\\ 
\vc{H}_{\ind{6}}& = H_{\ind{1}} \left(\, -\cos\theta_{\ind{1}}\csphi
                  ,\,\pm \cos\theta_{\ind{1}}\snphi ,\,  \sin\theta_{\ind{1}} \, \right) 
  \end{aligned}
\label{eq010046}  
\end{equation}
where the upper and lower signs correspond to the mirror-symmetric configuration
or to the twofold rotationally symmetric
configuration, respectively.

Let $\vc{\kappa}_{\ind{n}}$ ($n=1,2,\ldots,6$) be the wave vector
after the $n$th Bragg reflection. The wave vectors of x-rays reflected
from crystals with diffraction vectors $\vc H_{\ind{n}}$ are related
by momentum conservation,
\begin{equation}
  \vc{\kappa}_{\ind{n}}
  =
  \vc{\kappa}_{\ind{n-1}}
  +
  \vc{\tilde H}_{\ind{n}},
  \qquad
  \vc{\tilde H}_{\ind{n}}
  =
  \vc H_{\ind{n}}\left(1+w_{\ind{Hn}}\right).
  \label{eq0110}
\end{equation}
Elastic scattering requires conservation of photon energy,
\begin{equation}
  |\vc{\kappa}_{\ind{n}}|=\kappa,
  \qquad
  \kappa=\frac{E}{\hbar c},
  \label{eq0115}
\end{equation}
and leads to Bragg's law,
\begin{equation}
  2\kappa\sin\theta_{\ind{n}}
  =
  \tilde H_{\ind{n}}
  =
  H_{\ind{n}}\left(1+w_{\ind{Hn}}\right).
  \label{eq0130}
\end{equation}
Here $\theta_{\ind{n}}$ is the glancing angle of incidence with respect
to the reflecting atomic planes at the center of the Bragg-reflection
peak, i.e., the Bragg angle. The quantity $w_{\ind{Hn}}$ is the
Bragg-law correction due to refraction at the vacuum--crystal
interface. In the low-photoabsorption approximation, valid for
crystals such as diamond or Si, $w_{\ind{Hn}}$ is invariant, i.e.,
independent of $E$ and $\theta_{\ind{n}}$, for a given Bragg
reflection with diffraction vector $\vc H_{\ind{n}}$~\cite{Shvydko-SB}.
This statement applies here to symmetric Bragg reflections, for which
the reflecting atomic planes are parallel to the crystal entrance
surface.

\begin{figure}[t!]
\setlength{\unitlength}{\textwidth}
\includegraphics[width=0.50\textwidth]{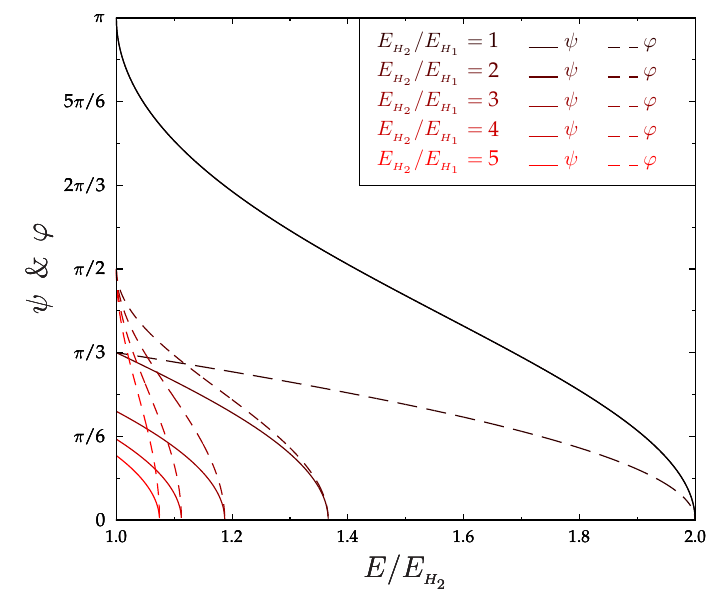}
\caption{Crystal rotation angles $\psi$ and $\varphi$ as a function of
  $E/E_{\ind{H_2}}$---the photon energy $E$ normalized to the
  backscattering energy $E_{\ind{H_2}}$---for different values of
  $E_{\ind{H_2}}/E_{\ind{H_1}}$, as given by
  Eqs.~\eqref{eq0012}--\eqref{eq0015}. No configurations are possible
  if $E_{\ind{H_2}}/E_{\ind{H_1}} < 1$. The tuning range is maximized
  when $E_{\ind{H_2}}=E_{\ind{H_1}}$. }
\label{fig2} 
\end{figure}

Equivalently, Bragg's law for a photon of energy $E$ can be written as
\begin{equation}
  E\sin\theta_{\ind{n}}=E_{\ind{Hn}},
  \qquad
  E_{\ind{Hn}}
  =
  \frac{\hbar c}{2}\,
  H_{\ind{n}}\left(1+w_{\ind{Hn}}\right),
  \label{eq0135}
\end{equation}
where $E_{\ind{Hn}}$ is the photon energy corresponding to exact
backscattering for the Bragg reflection 
with diffraction-vector magnitude $H_{\ind{n}}$.

We require that, after the third reflection the wave vector 
\begin{equation}
\vc{\kappa}_{\ind{3}}  = \vc{\kappa}_{\ind{0}} +\vc{\tilde{H}}_{\ind{1}} +\vc{\tilde{H}}_{\ind{2}} +\vc{\tilde{H}}_{\ind{3}}
\label{eq0140}
\end{equation}
becomes
antiparallel to that of the incident x-rays:
$\vc{\kappa}_{\ind{3}}  = -\vc{\kappa}_{\ind{0}}$. This condition leads to the system of equations
\begin{align}
-2\tilde{H}_{\ind{1}}\, \cos\theta_{\ind{1}}\snphi\, +\, \tilde{H}_{\ind{2}}\, \sin\psi & = 0\\ 
2\tilde{H}_{\ind{1}}\, \sin\theta_{\ind{1}}\, + \, \tilde{H}_{\ind{2}}\, \cos\psi & = 2 \kappa,
\end{align}
from which the crystal rotation angles $\psi$ and $\varphi$ can be
determined. Combining these equations with Bragg's law,
Eq.~\eqref{eq0130}, yields the following expressions for $\psi$ and
$\varphi$:
\begin{equation} 
\cos\psi = \frac{\cos 2\theta_{\ind{1}}}{\sin \theta_{\ind{2}}}, \hspace{0.5cm}
\cos \varphi = \frac{\cos \theta_{\ind{2}}}{\sin 2\theta_{\ind{1}}}  
\label{eq0010}
\end{equation}

The equivalent solutions for $\psi$ and $\varphi$ as a function of
photon energy $E$ are
\begin{align}
\label{eq0012}
  \cos\psi & = \frac{1-2(E_{\ind{H_1}}/E)^2}{E_{\ind{H_2}}/E},\\
   \cos\varphi\,  & =  \frac{E}{2E_{\ind{H_1}}}\frac{\sqrt{1-(E_{\ind{H_2}}/E)^2}}{\sqrt{1-(E_{\ind{H_1}}/E)^2}}. 
\label{eq0015}
\end{align}

\begin{figure}[t!]
\includegraphics[width=0.50\textwidth]{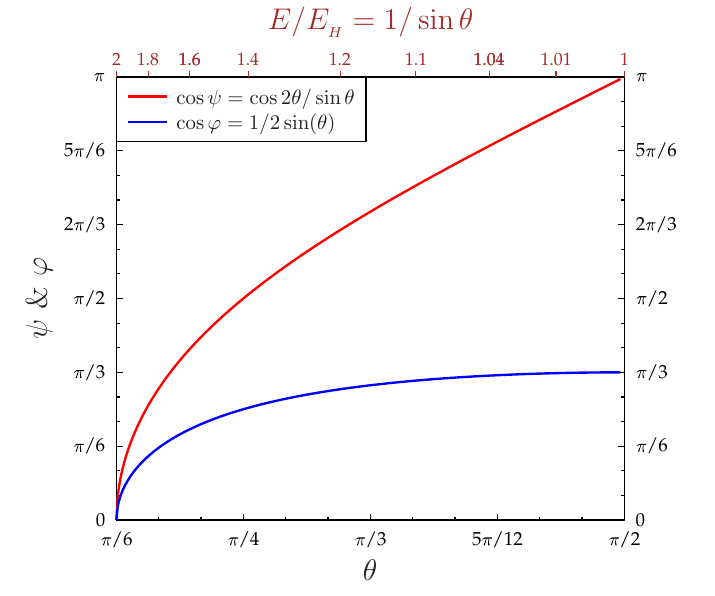}  
\caption{Graphical representation of Eqs.~\eqref{eq0020} for the
crystal rotation angles $\psi$ and $\varphi$ as functions of the Bragg
angle $\theta$, assuming that all reflections in the cavity are
identical. The upper energy scale is derived from Bragg's law,
Eq.~\eqref{eq0135}, and represents the ultimate cavity tuning range
from $E=E_{\ind{H}}$ to $2E_{\ind{H}}$ for a given reflection with
diffraction vector $\vc H$.}
\label{fig3} 
\end{figure}

\begin{figure*}[t]
\setlength{\unitlength}{\textwidth}
\includegraphics[width=0.99\textwidth]{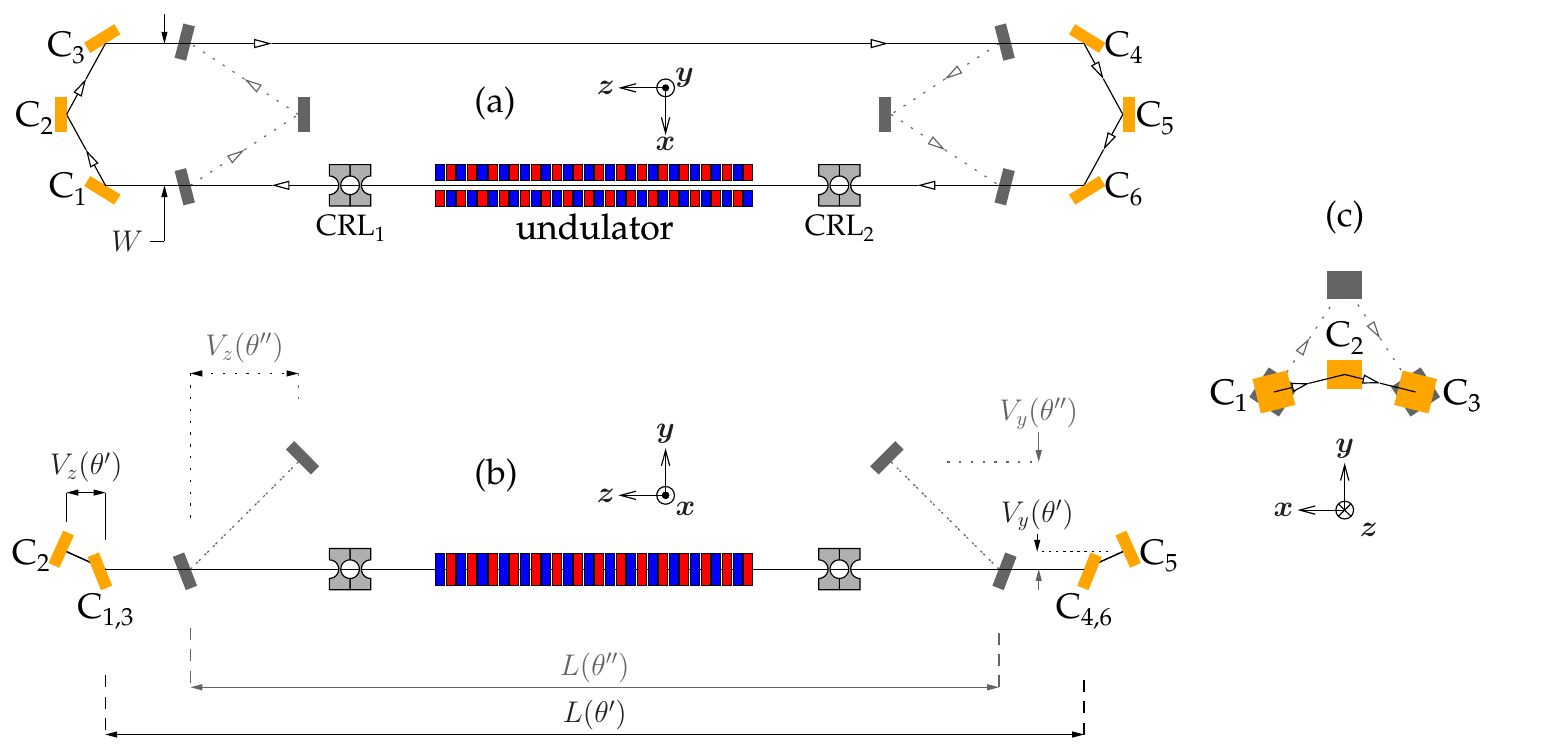}
\caption{Kinematics of the optical elements in the six-crystal
  noncoplanar x-ray cavity during photon-energy tuning, shown for two
  Bragg-angle settings: $\theta^{\prime}=31^{\circ}$ colored crystal
  elements and $\theta^{\prime\prime}=65^{\circ}$ superimposed gray
  elements.  (a) Top view.  (b,c) Front and side views of the
  cavity in the mirror-symmetric configuration.  Crystals C$_{\ind{2}}$ and
  C$_{\ind{5}}$ are translated within the backscattering units along
  the $z$-axis by $V_{\ind{z}}(\theta)$, while their vertical
  positions are adjusted to $V_{\ind{y}}(\theta)$, as given by
  Eq.~\eqref{eq080}, with the cavity width $W$ kept fixed. The x-ray
  round-trip length $\ell$ is kept constant by adjusting the
  inter-unit distance $L(\theta)$ according to Eq.~\eqref{eq082}.}
\label{fig4}
\end{figure*}

Figure~\ref{fig2} shows a graphical representation of these equations
as a function of $E/E_{\ind{H_2}}$---the photon energy $E$
normalized to the backscattering energy $E_{\ind{H_2}}$---for
different values of $E_{\ind{H_2}}/E_{\ind{H_1}}$. Two important
observations can be made. First, no cavity configurations are possible
if $E_{\ind{H_2}}/E_{\ind{H_1}} < 1$. Second, the tuning range is
maximized when $E_{\ind{H_2}}=E_{\ind{H_1}}$, i.e., when all
reflections are identical. In this case, the tuning range is from
$E = E_{\ind{H}}$ to $E = 2E_{\ind{H}}$,
corresponding to a 100\% tunability. In the remainder of this paper,
we analyze the cavity with identical reflections.

In this case
$\theta_{\ind{1}}=\theta_{\ind{2}}=\theta$, and Eq.~\eqref{eq0010}
simplifies to
\begin{equation} \cos\psi = \frac{\cos 2\theta}{\sin \theta},
  \hspace{1cm} \cos\varphi =
  \frac{1}{2\sin\theta}
  \label{eq0020}
\end{equation}
These relations define the accessible range
$\pi/6 \le \theta \le \pi/2$, which begins with the coplanar
configuration ($\varphi=0$, $\psi=0$) at $\theta=\pi/6$ and determines
the photon energy tuning range. Equations~\eqref{eq0020} are shown
graphically in Fig.~\ref{fig3}. It illustrates, in particular, that
this cavity has a 100\% tuning range, from $E=E_{\ind{H}}$ at
$\theta=\pi/2$ to $E=2E_{\ind{H}}$ at $\theta=\pi/6$, when the cavity
becomes coplanar.

\subsection{Cavity kinematics}
\label{kinematics}

Photon-energy tuning requires coordinated changes of the Bragg angle
$\theta$, the crystal rotation angles $\varphi$ and $\psi$, and the
crystal positions, so that the x-ray trajectory remains closed.  The
separation between the two backscattering units must also be adjusted
to keep the round-trip cavity length $\ell$ fixed and synchronized
with the period $\tau=\ell/v$ of the incoming relativistic electron
bunches.  Here $v\simeq c(1-1/2\gamma^2)$, where $c$ is the speed of
light and $\gamma\gg 1$ is the Lorentz factor.

Figure~\ref{fig4} illustrates two cavity configurations, corresponding
to Bragg angles $\theta^{\prime}=31^{\circ}$ and
$\theta^{\prime\prime}=65^{\circ}$, and shows the required motion of
the optical elements during photon-energy tuning.

We assume that the cavity width $W$ and round-trip length $\ell$ are
fixed.  The relevant quantities are the separations between the
backscattering units,
C$_{\ind{1}}$C$_{\ind{6}}$ =
C$_{\ind{3}}$C$_{\ind{4}}$ = $L(\theta)$,
and the intra-unit crystal separations,
C$_{\ind{1}}$C$_{\ind{2}}$ =
C$_{\ind{2}}$C$_{\ind{3}}$ =
C$_{\ind{4}}$C$_{\ind{5}}$ =
C$_{\ind{5}}$C$_{\ind{6}}$ = $V(\theta)$.
Their projections on the $z$- and $y$-axes are denoted by
$V_{\ind{z}}(\theta)$ and $V_{\ind{y}}(\theta)$, respectively.

The radius vector from C$_{\ind{1}}$ to C$_{\ind{2}}$ is parallel to
the wavevector $\vc{\kappa}_{\ind{1}}$ given by Eq.~\eqref{eq0A100}.
in Appendix~\ref{polarizationbasis}. Using this relation
together with Eq.~\eqref{eq0020}, one obtains
\begin{equation}
\begin{array}{ll}
V(\theta) &= \displaystyle \frac{W}{2\cos\theta}, \\[10pt]
V_{\ind{y}}(\theta) &= W\sin\theta\sin\varphi, \\[6pt]
V_{\ind{z}}(\theta) &= \displaystyle \frac{W\cos 2\theta}{2\cos\theta}.
\end{array}
\label{eq080}
\end{equation}

The optical path through each backscattering unit,
C$_{\ind{1}}$--C$_{\ind{3}}$ or C$_{\ind{4}}$--C$_{\ind{6}}$, is
$2V$.  Since the round-trip length is
\begin{equation}
\ell = 4V + 2L,
\end{equation}
a change in $V(\theta)$ must be compensated by changing $L(\theta)$.
Relative to the coplanar configuration at $\theta=\pi/6$, the required
translation of each backscattering unit is
\begin{equation}
\Delta L
= [L(\pi/6)-L(\theta)]/2
= \frac{W}{2}\left(\frac{1}{\cos\theta}-\frac{2}{\sqrt{3}}\right).
\label{eq082}
\end{equation}

\begin{figure}[t!]
\includegraphics[width=0.50\textwidth]{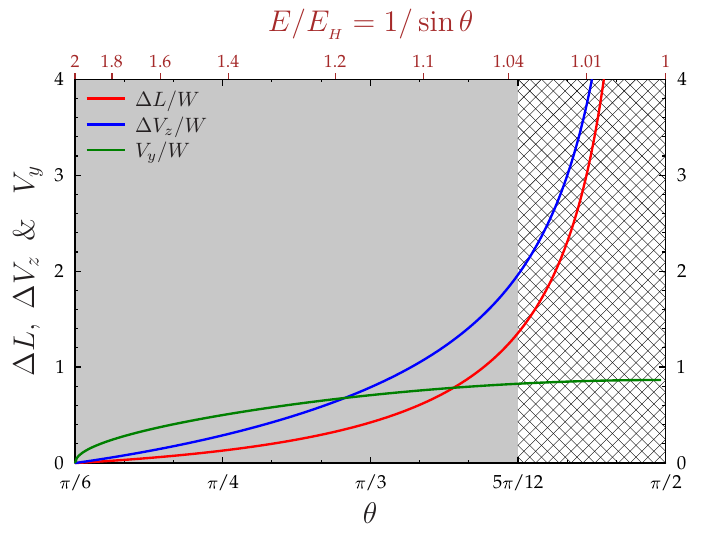}
  \caption{Spatial adjustments of the cavity during photon-energy
  tuning.  The plot shows the backscattering-unit translation
  $\Delta L$ from Eq.~\eqref{eq082}, together with the $z$- and
  $y$-axis projections of the C$_{\ind{2}}$ and C$_{\ind{5}}$
  displacements, $\Delta V_{\ind{z}}(\theta)$ and
  $V_{\ind{y}}(\theta)$, from Eqs.~\eqref{eq084} and \eqref{eq080}.
  All quantities are normalized to the cavity width $W$. The
  cross-hatched region, corresponding to large Bragg angles
  $75^{\circ} \lesssim \theta \leq 90^{\circ}$, requires large
  longitudinal dimensions of the backscattering units, as determined
  by $\Delta V_{\ind{z}}$, together with large longitudinal
  translations $\Delta L$.}
\label{fig7}
\end{figure}

At the same time, crystals C$_{\ind{2}}$ and C$_{\ind{5}}$ must be
translated in the backscattering units along the $z$-axis by
\begin{equation}
\Delta V_{\ind{z}}
= V_{\ind{z}}(\pi/6)-V_{\ind{z}}(\theta)
= \frac{W}{2}\left(
\frac{1}{\sqrt{3}}-\frac{\cos 2\theta}{\cos\theta}
\right),
\label{eq084}
\end{equation}
while their vertical positions are adjusted according to
$V_{\ind{y}}(\theta)$.  Equations~\eqref{eq080}--\eqref{eq084} therefore
define the spatial kinematics of the cavity crystals.  These
dependencies are graphically represented in Fig.~\ref{fig7}.

The cross-hatched region in Fig.~\ref{fig7}, corresponding to large
Bragg angles $75^{\circ} \lesssim \theta \leq 90^{\circ}$, requires
large longitudinal dimensions of the backscattering units, as
determined by $\Delta V_{\ind{z}}$, together with large longitudinal
translations $\Delta L$.  Since the photon energy increases by only 3.5\%, from
$E_{\ind{H}}$ to approximately $1.035E_{\ind{H}}$, over this range,
the corresponding region may be excluded from practical operation.

\section{Cavity reflectivity}

In this section, we study the reflectivity of the noncoplanar
six-crystal cavity as functions of incident photon energy and
polarization, and examine how the spectral range of high reflectivity
and the polarization state of the reflected photons vary with crystal
Bragg angle. We consider a well-collimated beam with negligible
divergence. The cavity reflectivity is determined by the Bragg
reflectivity of the individual crystals, as given by the well-known
expressions of the dynamical theory of x-ray Bragg diffraction in
crystals, comprehensively reviewed in \cite{Zach,Pinsker,Authier,Shvydko-SB}.

\subsection{Bragg reflectivity of a single crystal}

We consider a particular case of Bragg reflection of a plane
monochromatic x-ray wave from a thick centrosymmetric crystal (e.g.,
diamond or silicon) in symmetric scattering geometry.
For simplicity, we assume the non-absorbing crystal approximation,
with vanishing imaginary parts of the Fourier components of
the crystal electric susceptibility $\chi_{\ind{0}}$ and
$\chi_{\ind{H}}$, which is a good approximation for diamond and
reasonable for silicon crystals.

For each pair of incident and reflected waves with wave vectors
$\vc{\kappa}_{\ind{n-1}}$ and
$\vc{\kappa}_{\ind{n}}=\vc{\kappa}_{\ind{n-1}}+\vc{\tilde{H}}_{\ind{n}}$,
respectively, associated with the diffraction vector
$\vc{\tilde{H}}_{\ind{n}}$ of crystal C$_{\ind{n}}$ (here
$n=1,2,..,6$) we define a set of mutually orthogonal linear
polarization unit vectors as
\begin{equation}
\vc{\sigma}_{\ind{nm}}=\frac{\vc{\kappa}_{\ind{n}}\times
    \vc{\kappa}_{\ind{n-1}}}{|\vc{\kappa}_{\ind{n}}\times \vc{\kappa}_{\ind{n-1}}|}, \hspace{0.3cm}
\vc{\pi}_{\ind{nm}}=\frac{\vc{\kappa}_{\ind{n-1+m}}\times
  \vc{\sigma}_{\ind{nm}}}{\kappa},
\label{eq0060}
\end{equation}
where $m=0$ and $m=1$ correspond to the incident and reflected waves, respectively.
The unit vectors of the $\sigma$-polarization components are
perpendicular to the diffraction plane
($\vc{\kappa}_{\ind{n-1}},\vc{\kappa}_{\ind{n}})$, while
the unit vectors of the $\pi$-polarization components lie in the
diffraction plane.

The corresponding reflectivity amplitudes $R^{s\tilde{s}}$ for an
incident x-ray wave in the linear polarization state
$\tilde{s}=\sigma$ or $\tilde{s}=\pi$ are given by
\begin{equation}
R^{s\tilde{s}}= -\sgn(P^s) 
\left[  -\frac{\eta}{|P^s|}
  \pm \sqrt{\left(\frac{\eta}{|P^s|}\right)^2-1}
\,  \right] \, \delta^{s\tilde{s}},
\label{eq357}
\end{equation}
where 
\begin{equation}
P^s=
\begin{cases}
  1, & \text{if $s=\sigma$},\\
 \cos 2\theta , & \text{if $s=\pi$}.
\end{cases}
\label{eq359}
\end{equation}
is the polarization factor, and $\eta$ is the reduced deviation
parameter from Bragg's condition. 

In particular, for a fixed glancing angle of incidence $\theta$,
$\eta$ relates to relative photon-energy deviation $\delta E$
from the value $E$ defined by
Bragg's law Eq.~\eqref{eq0135} as
\begin{equation}
\eta=-2\frac{\delta E}{\epsilon_{\ind{H}}\,E}
\label{eq0082}
\end{equation}
where $\epsilon_{\ind{H}} = -{\chi_{\ind{H}}}/{\sin^2\theta}$ is along
with $w_{\ind{H}} = -{\chi_{\ind{0}}}/{2\sin^2\theta}$ another
Bragg-reflection invariant corresponding to its relative spectral
width \cite{Shvydko-SB}.  It is also directly related to one more
important invariant -- the extinction length,
\begin{equation}
\Lambda_{\ind{H}}=\frac{2}{H \epsilon_{\ind{H}}},
\label{eq00822}
\end{equation}
which characterizes the propagation length of x-rays in a crystal
under Bragg diffraction. Numerical values of these Bragg-reflection
invariants in Si crystals are given in \cite{Shvydko-SB}, and those
for diamond crystals are reported in \cite{SL12}. Using
Eq.\eqref{eq00822} for the extinction length, the expression for the
deviation parameter in Eq.\eqref{eq0082} can be rewritten as
\begin{equation}
\eta=-\frac{\delta E\,\,\timex}{\hbar}, 
\label{eq00824}
\end{equation}
where 
\begin{equation}
\timex=\frac{2\Lambda_{\ind{H}}\sin\theta}{c}.
\label{eq00826}
\end{equation}
is the characteristic time delay of x-rays associated with Bragg diffraction in
the crystal in the $\sigma$-polarization state arising from the finite extinction length \cite{Shvydko-SB,LSKF11,SL12}.

According to Eq.~\eqref{eq357}, the $\sigma$- and $\pi$-polarization states determined by Eq.~\eqref{eq0060}
are eigenpolarization states with no coupling between them.

The sign before the square root in Eq.~\eqref{eq357} is chosen
positive for $\eta>-1$ and negative otherwise.  Importantly, the sign
of the reflection amplitude for the $\pi$ polarization component flips
at $\theta=\pi/4$. This leads to important consequences for
polarization coupling in noncoplanar multi-crystal arrangements
discussed in the paper.

In the range $|\eta/|P^s|\leq 1$ Eq.~\eqref{eq357} becomes
\begin{equation}
R^{ss}= -\sgn({P^s})
\left[  -\frac{\eta}{|P^s|}
  +i \sqrt{1-\left(\frac{\eta}{P^s}\right)^2}
\,  \right] 
\label{eq358}
\end{equation}
resulting in $|R^{ss}|^2=1$, i.e., in total reflection with the center
at $\eta=0$.  The region of total Bragg reflection for
$\pi$-polarization component $|\eta|\leq |P^s|$ is always smaller than
the total-reflection region $|\eta|\leq 1$ for $\sigma$-polarization.
In systems with the polarization mixing such as the considered here
noncoplanar cavity, the former determines the region of high
reflectivity of the whole system.

The reflection amplitude $R^{\pi\pi}$ for $\pi$-polarization has a
singularity at $\theta = 45^\circ$ because $P^\pi = 0$, causing the
reflection width to vanish as well. Since this configuration is not
practical, it is not considered further here.

Equation~\eqref{eq358} can be presented in an equivalent form
\begin{equation}
R^{ss}\,=\, -i\sgn({P^s})\, \exp(i\,\phi^s),
\label{eq3591}  
\end{equation}
where
\begin{equation}
\sin\phi^s\,=\,\frac{\eta}{|P^s|}.
\label{eq3592}
\end{equation}
Here, the phase $\phi^s$ changes from $-\pi/2$ to $+\pi/2$ within the
total-reflection region $(-1<\eta/|P^s|<1)$. Since the total-reflection
widths differ for the two polarization states, the relative phase
between the polarization components changes upon Bragg reflection.  As
a result, their superposition changes, leading to a modification of
the polarization state of the reflected radiation. Consequently,
radiation that is initially linearly polarized generally becomes
elliptically polarized after successive Bragg reflections in the
cavity as detailed below.

\subsection{Bragg reflectivity of the  six-crystal cavity}

The electric-field amplitude of the photon with arbitrary polarization
incident on the first crystal can be written as
\begin{equation}
  \vc E_{\indrm{i}}  =  E_{\indrm{i}}^{\sigma}\esi  +  E_{\indrm{i}}^{\pi}\epi ,
  \label{eq0510}
\end{equation}
where $E_{\indrm{i}}^{\sigma}$ and $E_{\indrm{i}}^{\pi}$ are the
amplitudes of the two mutually orthogonal components in the
geometrically defined intrinsic polarization basis
\begin{equation}
  \esi=\hat{\vc y},
  \qquad
  \epi=-\hat{\vc x}.
  \label{eq051}
\end{equation}
Here $\epi$ lies in the principal cavity plane $(x,z)$, whereas
$\esi$ is perpendicular to it. We first calculate the cavity
reflectivity in this intrinsic polarization basis.

\subsubsection{First reflection}

The intrinsic polarization basis $\{\esi,\epi\}$ is related to the
linear polarization basis
$\{\vc{\sigma}_{\ind{10}},\vc{\pi}_{\ind{10}}\}$, defined with respect
to the diffraction plane of the first crystal, by rotating the
principal cavity plane $(x,z)$ about the $\hat{\vc z}$ axis by an
angle $\varphi$, with positive rotation defined as counterclockwise:
\begin{equation}
\left(\!\! \begin{array}{r}
 \vc{\sigma}_{\ind{10}} \\ \vc{\pi}_{\ind{10}}
\end{array} \right)
= \hat{U}(-\varphi)
\left(\!\! \begin{array}{r}
 \esi \\ \epi
\end{array} \right)
\label{eq048}
\end{equation}
where
\begin{equation}
\hat{U}(\varphi)=
  \left( \begin{array}{cc}
    \cos\varphi &   \sin\varphi  \\
      -\sin\varphi &  \cos\varphi  \\
  \end{array} \right)
\label{eq047}
\end{equation}
is the rotation matrix. See also Eq.~\eqref{eq0A2020} of Appendix~\ref{polarizationbasis}.
The reflection amplitudes of the first crystal
in the polarization basis $(\vc{\sigma}_{\ind{11}},\vc{\pi}_{\ind{11}})$ is then given by
\begin{equation}
\left(\!\! \begin{array}{r}
 E_{\ind{1}}^{\sigma} \\ E_{\ind{1}}^{\pi} 
\end{array} \right)
=
    \, \hat{R}_{\ind{1}}^{\,\prime} 
\left(\!\! \begin{array}{r}
 E_{\indrm{i}}^{\sigma} \\ E_{\indrm{i}}^{\pi} 
\end{array} \right),
\qquad \hat{R}_{\ind{1}}^{\,\prime}\, =   \, \hat{R}(\eta_{\ind{1}}) \, \hat{U}(-\varphi),
\label{eq045}
\end{equation}
where
\begin{equation}
  \hat{R}(\eta_{\ind{1}})  =
  \left[ \begin{array}{cc}
    R^{\sigma\sigma}(\eta_{\ind{1}})  &   0   \\
      0 &  R^{\pi\pi}(\eta_{\ind{1}}) 
  \end{array} \right] 
\label{eq046}
\end{equation}
is the single crystal reflection matrix, whose amplitudes are given either
by Eq.~\eqref{eq357} within the aforementioned approximation or by the
more accurate expressions of \cite{Zach,Pinsker,Authier,Shvydko-SB}.
In the common total-reflection region, $|\eta/P^\pi|<1$, the matrix
$\hat{R}(\eta_{\ind{1}})$ can also be expressed using Eqs.~\eqref{eq3591}--\eqref{eq3592} as
\begin{equation}
  \hat{R}(\eta_{\ind{1}})  =
  -i\left[ \begin{array}{cc}
    \exp(i\,\phi^\sigma) &   0   \\
      0 &  \sgn({P^\pi})\, \exp(i\,\phi^\pi) 
  \end{array} \right] 
\label{eq0462}
\end{equation}
where $\sin\phi^s={\eta_{\ind{1}}}/{|P^s|}$.

\subsubsection{Two reflections}

As shown in Eq.~\eqref{eq0A2010} of Appendix~\ref{polarizationbasis},  the polarization basis
$(\vc{\sigma}_{\ind{11}},\vc{\pi}_{\ind{11}})$ of the photon reflected
from the first crystal 
is transformed to the eigenpolarization basis of the second crystal  $(\vc{\sigma}_{\ind{20}},\vc{\pi}_{\ind{20}})$ as
\[
\left(\!\! \begin{array}{r}
 \vc{\sigma}_{\ind{20}} \\ \vc{\pi}_{\ind{20}}
\end{array} \right)
= \hat{U}(2\varphi)
\left(\!\! \begin{array}{r}
 \vc{\sigma}_{\ind{11}} \\ \vc{\pi}_{\ind{11}}
\end{array} \right)
\]
As a result, the reflection amplitudes from the second crystal
in the polarization basis $(\vc{\sigma}_{\ind{21}},\vc{\pi}_{\ind{21}})$ is given by
\begin{equation}
\left(\!\! \begin{array}{r}
 E_{\ind{2}}^{\sigma} \\ E_{\ind{2}}^{\pi} 
\end{array} \right)
= \, \hat{R}_{\ind{2}}^{\,\prime}
\left(\!\! \begin{array}{r}
 E_{\indrm{i}}^{\sigma} \\ E_{\indrm{i}}^{\pi} 
\end{array} \right),
\qquad \hat{R}_{\ind{2}}^{\,\prime}=\hat{R}(\eta_{\ind{2}})\, \hat{U}(2\varphi)    \,\hat{R}_{\ind{1}}^{\,\prime}.
\label{eq049}
\end{equation}

\subsubsection{Three subsequent reflections}

Similarly, the reflection amplitudes from the third crystal
in the polarization basis $(\vc{\sigma}_{\ind{31}},\vc{\pi}_{\ind{31}})$ are given by
\begin{equation}
\left(\!\! \begin{array}{r}
 E_{\ind{3}}^{\sigma} \\ E_{\ind{3}}^{\pi} 
\end{array} \right)
= \hat{R}_{\ind{3}}^{\,\prime}
\left(\!\! \begin{array}{r}
 E_{\indrm{i}}^{\sigma} \\ E_{\indrm{i}}^{\pi} 
\end{array} \right)
\qquad
\hat{R}_{\ind{3}}^{\,\prime}\,=\, \hat{R}(\eta_{\ind{3}})\hat{U}(-2\varphi) \hat{R}_{\ind{2}}^{\,\prime}
\label{eq050}
\end{equation}

To express the reflectivity amplitudes after the three successive
reflections in the initial polarization basis of Eq.~\eqref{eq051}, an
additional rotation is required to bring the diffraction plane
($\vc{\kappa}_{\ind{2}},\vc{\kappa}_{\ind{3}})$ into coincidence with
the principal cavity plane $(x,z)$. The resulting cumulative
reflection matrix becomes
\[\hat{R}_{\ind{3}}\,=\, \hat{U}(\varphi) \hat{R}_{\ind{3}}^{\,\prime} \]
or
\begin{equation}
  \hat{R}_{\ind{3}} = \hat{U}(\varphi)\,  \hat{R}(\eta_{\ind{3}})\, \hat{U}(-2\varphi)\, \hat{R}(\eta_{\ind{2}})\, \hat{U}(2\varphi)\, \hat{R}(\eta_{\ind{1}})\, \hat{U}(-\varphi) 
\label{eq0305}
\end{equation}
For the case of all three identical crystals and assuming
$\eta_{\ind{1}}=\eta_{\ind{2}}=\eta_{\ind{3}}=\eta$ valid for
x-ray beam   with negligible angular divergence the equation
can be presented as a product of three single-reflection matrices
\begin{equation}
\hat{R}_{\ind{3}} = \hat{Q}(-\varphi)\,  \hat{Q}(\varphi)\, \hat{Q}(-\varphi) 
\label{eq0300} 
\end{equation}
where the single-reflection matrix is
\[
\hat{Q}(\varphi) = \hat{U}(-\varphi)\, \hat{R}(\eta)\, \hat{U}(\varphi).  
\]
Expressing $R^{\sigma\sigma}=v+u$ and $R^{\pi\pi}=v-u$ via the
symmetric $v$ and antisymmetric $u$ co-polar amplitudes
\begin{equation}
 v=\frac{R^{\sigma\sigma}+R^{\pi\pi}}{2}, \text{~~~~and} \qquad u=\frac{R^{\sigma\sigma}-R^{\pi\pi}}{2}. 
\label{eq043}
\end{equation}
the single-reflection matrix becomes
\[
  \hat{Q}(\varphi) =
  \begin{pmatrix}
 v+u \cos 2\varphi & u \sin 2 \varphi \\
 u \sin 2 \varphi & v-u \cos 2\varphi 
\end{pmatrix}
,
\]
and the three-reflection propagation matrix 
\begin{equation}
\hat R_{\ind{3}}=
\begin{pmatrix}
  R_{\ind{3}}^{\sigma\sigma}  & R_{\ind{3}}^{\sigma\pi} \\
R_{\ind{3}}^{\pi\sigma}  & R_{\ind{3}}^{\pi\pi} 
\end{pmatrix}
=
\begin{pmatrix}
a & b\\
b & d
\end{pmatrix}
\label{eq053}
\end{equation}
where
\begin{equation}
  \begin{split}
a=& v \left(u^2+v^2+3 u v \cos 2\varphi + 2 u^2 \cos
    4\varphi\right)+u^3  \cos 6\varphi\\
  b=& -u \left(v^2 \sin 2 \varphi + u^2 \sin 6 \varphi\right) \\ 
d=& v \left(u^2+v^2-3 u v \cos 2\varphi+2 u^2 \cos 4\varphi\right)-u^3
    \cos 6\varphi .
  \end{split}
\label{eq052}            
\end{equation}
In particular, for $\varphi=0$, which corresponds to a coplanar cavity with $\theta=\pi/6$, Eq.~\eqref{eq052} reduces to the anticipated result
\begin{equation}
\begin{array}{lll}
R_{\ind{3}}^{\sigma\sigma}&=a=(v+u)^3&=(R^{\sigma\sigma})^3,\\
R_{\ind{3}}^{\pi\sigma}=R_{\ind{3}}^{\sigma\pi}&=b&=0,\\
R_{\ind{3}}^{\pi\pi}&=d=(v-u)^3&=(R^{\pi\pi})^3,
\end{array}
\label{eq0523}
\end{equation}
with no polarization coupling. The three-crystal reflectivity amplitudes are simply equal to the cubes of the corresponding single-crystal reflectivity amplitudes, Eq.~\eqref{eq357}, for each of the two orthogonal linear polarization components.

In another limiting case, corresponding to nearly exact
backscattering in a single reflection, $\theta\rightarrow\pi/2$ and
$\varphi\rightarrow\pi/3$. According to Eq.~\eqref{eq359}, the
polarization factor satisfies $P^\pi\simeq-1$, i.e., it has
approximately the same magnitude but the opposite sign as
$P^\sigma=1$. Consequently, the single-crystal reflectivity
amplitudes, Eq.~\eqref{eq357}, satisfy $R^{\sigma\sigma}\simeq
-R^{\pi\pi}$. It then follows from Eq.~\eqref{eq043} that $v\simeq0$
and $u\simeq R^{\sigma\sigma}$. Substituting these relations into
Eq.~\eqref{eq052} yields also expected result
\begin{equation}
\begin{array}{lll}
R_{\ind{3}}^{\sigma\sigma}&=a\simeq u^3&=(R^{\sigma\sigma})^3,\\
R_{\ind{3}}^{\pi\sigma}=R_{\ind{3}}^{\sigma\pi}&=b\simeq -({\sqrt{3}}/{2})\,u\,v^2&\simeq0,\\
R_{\ind{3}}^{\pi\pi}&=d\simeq -u^3&=-(R^{\sigma\sigma})^3,
\end{array}
\label{eq0525}
\end{equation}
again with negligible polarization coupling. In this limit, the
three-crystal reflectivity amplitudes satisfy
$R_{\ind{3}}^{\sigma\sigma}\simeq -R_{\ind{3}}^{\pi\pi}$: they have
equal magnitudes, given by the cube of the 
single-crystal reflectivity amplitude $R^{\sigma\sigma}$, but opposite signs.

\subsubsection{Six subsequent reflections. Mirror-symmetric cavity.}

The propagation matrix for the next three successive reflections from
crystals C$_{\ind{4}}$-C$_{\ind{6}}$ in the second backscattering unit
in the mirror-symmetric configuration is given by the same matrix
$\hat R_{\ind{3}}$. Therefore, the propagation matrix for the
six-crystal cavity in the initial polarization basis is
$\hat R_{\ind{6m}}=\hat R_{\ind{3}}\, \hat R_{\ind{3}}$ and is given by
\begin{equation}
\hat R_{\ind{6m}}=
\begin{pmatrix}
R_{\ind{6m}}^{\sigma\sigma}  & R_{\ind{6m}}^{\sigma\pi} \\
R_{\ind{6m}}^{\pi\sigma}  & R_{\ind{6m}}^{\pi\pi} 
\end{pmatrix}
=
\begin{pmatrix}
a^2+b^2 & b(a+d)\\
b(a+d) & d^2+b^2
\end{pmatrix}
\label{eq054}
\end{equation}
using Eqs.~\eqref{eq053}-\eqref{eq052}.

In the coplanar cavity limit ($\theta=\pi/6$, $\varphi=0$),
Eqs.~\eqref{eq054}, together with Eq.~\eqref{eq0523}, reduce to
the anticipated solution
\begin{equation}
\begin{array}{lll}
R_{\ind{6m}}^{\sigma\sigma}&=(v+u)^6&=(R^{\sigma\sigma})^6,\\
R_{\ind{6m}}^{\pi\sigma}=R_{\ind{6m}}^{\sigma\pi}&=&0,\\
R_{\ind{6m}}^{\pi\pi}&=(v-u)^6&=(R^{\pi\pi})^6,
\end{array}
\label{eq0543}
\end{equation}
with no polarization coupling. The six-crystal reflectivity amplitudes are simply equal to the sixth powers of the corresponding single-crystal reflectivity amplitudes, $R^{\sigma\sigma}$ and $R^{\pi\pi}$.

In the backscattering limit ($\theta\rightarrow\pi/2$, $\varphi\rightarrow\pi/3$), Eqs.~\eqref{eq054}, together with Eq.~\eqref{eq0525}, reduce to the expected result
\begin{equation}
\begin{array}{lll}
R_{\ind{6m}}^{\sigma\sigma}&=u^6&=(R^{\sigma\sigma})^6,\\
R_{\ind{6m}}^{\pi\sigma}=R_{\ind{6m}}^{\sigma\pi}&=&0,\\
R_{\ind{6m}}^{\pi\pi}&=u^6&=(R^{\sigma\sigma})^6,
\end{array}
\label{eq0545}
\end{equation}
again with negligible polarization coupling. In this limit, the six-crystal reflectivity amplitudes become nearly identical,
$R_{\ind{6m}}^{\sigma\sigma}\simeq R_{\ind{6m}}^{\pi\pi}$,
and are equal to the sixth power of the single-crystal reflectivity amplitude $R^{\sigma\sigma}$.

\subsubsection{Six subsequent reflections.   Twofold rotationally  symmetric cavity.}

The propagation matrix for the three successive reflections in the
second backscattering unit, in the cavity with the twofold rotationally symmetric
configuration shown in Fig.~\ref{fig1}(d)–(e), is given by the
propagation matrix of the first backscattering unit in
Eqs.~\eqref{eq0305}–\eqref{eq053}, but with negative $ \varphi $:
\begin{equation}
\hat R_{\ind{3r}} =\hat R_{\ind{3}}(-\varphi) = \hat{Q}(\varphi)\,  \hat{Q}(-\varphi)\, \hat{Q}(\varphi) 
=
\begin{pmatrix}
a & -b\\
-b & d
\end{pmatrix}
.
\label{eq053r}
\end{equation}
Therefore, the propagation matrix for the
six-crystal cavity in the intrinsic polarization basis is
$\hat R_{\ind{6r}}=\hat R_{\ind{3r}}\ \hat R_{\ind{3}}$ and can be written as
\begin{equation}
\hat R_{\ind{6r}}=
\begin{pmatrix}
R_{\ind{6r}}^{\sigma\sigma}  & R_{\ind{6r}}^{\sigma\pi} \\
R_{\ind{6r}}^{\pi\sigma}  & R_{\ind{6r}}^{\pi\pi} 
\end{pmatrix}
=
\begin{pmatrix}
a^2-b^2 & -b(a-d)\\
b(a-d) & d^2-b^2
\end{pmatrix}
\label{eq054r}
\end{equation}
with the account of Eqs.~\eqref{eq053} and \eqref{eq053r}.

In the limiting coplanar and backscattering cases, the six-crystal
propagation matrix of the cavity in the twofold rotationally symmetric
configuration is given by the same equations, \eqref{eq0543} and
\eqref{eq0545}, as in the mirror-symmetric configuration.

The intermediate cases in both configurations are illustrated by the
numerical simulations presented in the next section.

\begin{turnpage}
  \begin{figure*}[p]
    \centering
\includegraphics[width=1.3\textwidth]{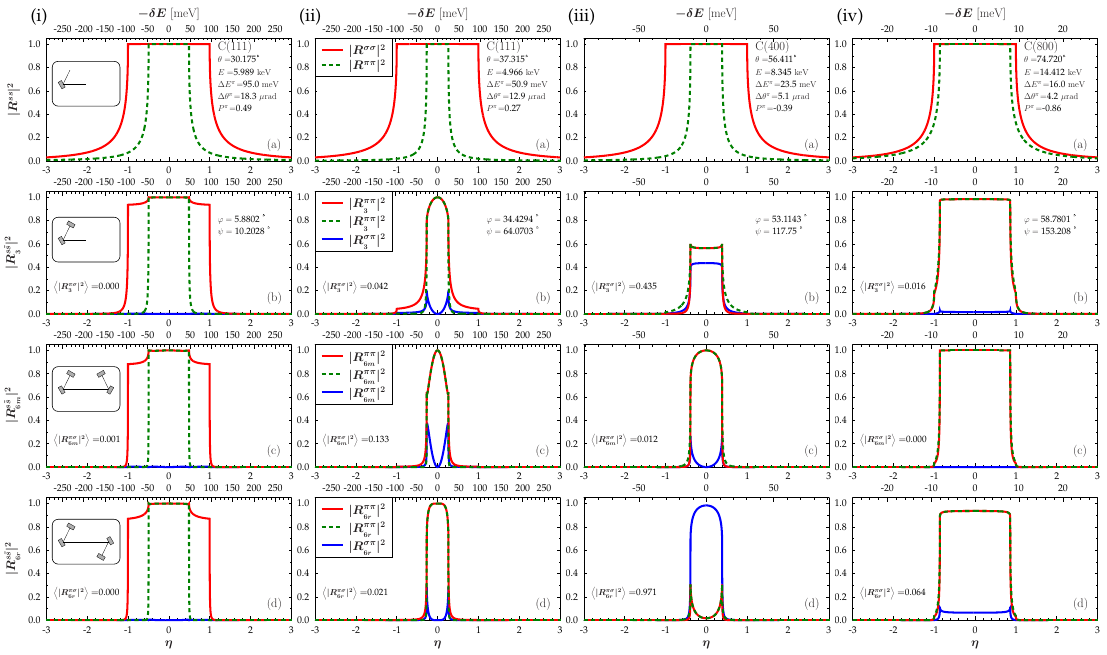}
\caption{Squared moduli $|R^{s\tilde{s}}|^2$ of the Bragg-reflection
polarization matrix elements in the intrinsic polarization basis
($s,\tilde{s}=\sigma,\pi$) as functions of the deviation parameter
$\eta$. Panels (i)--(iv) correspond to different Bragg angles
$\theta$. Each panel compares four cases of the Bragg reflectivity:
(a) a single crystal, calculated using Eq.~\eqref{eq357}; (b) the
cavity’s three-crystal backreflecting unit, calculated using
Eqs.~\eqref{eq043}–\eqref{eq052}; and (c) and (d) six successive
crystal reflections across the cavity’s two backreflecting units in
the mirror-symmetric and twofold rotationally symmetric
configurations, calculated using Eqs.~\eqref{eq054} and
\eqref{eq054r}, respectively.
The complementary relative
photon-energy scales $-\delta E$ are calculated using
Eq.~\eqref{eq0082} for selected diamond reflections with photon
energies of practical interest: (i) $E=5.989$~keV, corresponding to
the Cr $K$ edge~\cite{XDB}; (ii) $E=4.966$~keV, corresponding to the
Ti $K$ edge~\cite{XDB}; (iii) $E=8.345$~keV, corresponding to the
Ni $K$ edge~\cite{XDB}; and (iv) $E=14.4125$~keV, corresponding to
the $^{57}$Fe nuclear resonance~\cite{ShLJ00}.}
\label{fig51}
  \end{figure*}
\end{turnpage}

\subsubsection{Numerical simulations}
Figure~\ref{fig51} shows the squared moduli $ |R^{s\tilde{s}}|^2 $
of the Bragg-reflectivity polarization-matrix elements in the
intrinsic polarization basis $ (s,\tilde{s}=\sigma,\pi) $ as
functions of the deviation parameter $ \eta $ in panels
(i)–(iv). Each panel compares four cases of the Bragg reflectivity:
(a) a single crystal, calculated using Eq.~\eqref{eq357}; (b) the
cavity’s three-crystal backreflecting unit, calculated using
Eqs.~\eqref{eq043}–\eqref{eq052}; and (c) and (d) six successive
crystal reflections across the cavity’s two backreflecting units in
the mirror-symmetric and twofold rotationally symmetric
configurations, calculated using Eqs.~\eqref{eq054} and
\eqref{eq054r}, respectively.

The differences among the panels arise from the variation of the Bragg
angle $\theta$. Specifically, $\theta$ changes from $30.157^\circ$ in
the left panel (i), close to the coplanar-cavity case, to
$74.721^\circ$ in the right panel (iv), close to backscattering
conditions. Over this range, the polarization factor $P^\pi$ changes
from $P^\pi\simeq +0.49$ to $P^\pi\simeq -0.86$.

The three-crystal and six-crystal reflectivities are calculated in the
intrinsic polarization basis  $\{\esi,\epi\}$ defined in
Eq.~\eqref{eq051}. The complementary photon energy scale, $-\delta E$, is
calculated in each panel for a specific Bragg reflection in a diamond
crystal using Eq.~\eqref{eq0082}, with the Bragg-law photon energy
$E$ of interest related to the Bragg angle $\theta$ using
Eq.~\eqref{eq0135}. Finally, the admixture
$\langle|R^{\pi\sigma}_6|^2\rangle=\langle|R^{\sigma\pi}_6|^2\rangle$ of
the complementary polarization component after a complete round trip,
shown in each panel, is averaged over the region $|\eta|\leq 1/4$.
This quantity provides a measure of the cavity losses due to polarization mixing.

The comparison of the panels in Fig.~\ref{fig51} shows that in agreement
with the expectations, see Eqs.~\eqref{eq0543}-\eqref{eq0545}
discussed in preceding section, the polarization coupling is very weak
in the six-crystal cavity under conditions close to coplanar cavity --
panel (i) -- when $\theta \rightarrow 30^\circ$, and when close to
backscattering with $\theta \rightarrow 90^\circ$ -- panel
(iv).

As the Bragg angle $\theta$ moves away from $30^\circ$, for example
toward $37^\circ$ in panel (ii), the high-reflectivity region narrows
by a factor $1/|P^\pi|\simeq 4$ and the polarization coupling becomes
more pronounced. Nevertheless, at the center of the reflection range,
$\eta=0$, the coupling still vanishes both after three reflections and
after a full six-reflection round trip. Over the interval
$|\eta|\leq 1/4$, the average admixture of the orthogonal
polarization component remains moderate in the mirror-symmetric
cavity,
\[
\langle|R^{\pi\sigma}_{\ind{6m}}|^2\rangle
=
\langle|R^{\sigma\pi}_{\ind{6m}}|^2\rangle
\simeq 0.133,
\]
and is substantially smaller in the twofold rotationally symmetric
cavity,
\[
\langle|R^{\pi\sigma}_{\ind{6r}}|^2\rangle
=
\langle|R^{\sigma\pi}_{\ind{6r}}|^2\rangle
\simeq 0.021.
\]

The situation changes qualitatively once $\theta$ exceeds
$45^\circ$. As shown in panel (iii) for $\theta=56.41^\circ$, the
polarization coupling after three reflections no longer vanishes at
$\eta=0$; instead, the cross-polarization intensity
$|R_{\ind{3}}^{\pi\sigma}|^2$ becomes large. After six reflections,
however, corresponding to a complete round trip, the mirror-symmetric
cavity again suppresses the polarization coupling at the center of the
reflection range. Over the same interval $|\eta|\leq 1/4$, the
average admixture of the orthogonal polarization component is then
very small,
\[
\langle|R^{\pi\sigma}_{\ind{6m}}|^2\rangle
=
\langle|R^{\sigma\pi}_{\ind{6m}}|^2\rangle
\simeq 0.012.
\]
By contrast, in the twofold rotationally symmetric cavity the
polarization mixing is strongly enhanced and can approach unity.

\begin{figure}[t!]
\includegraphics[width=0.50\textwidth]{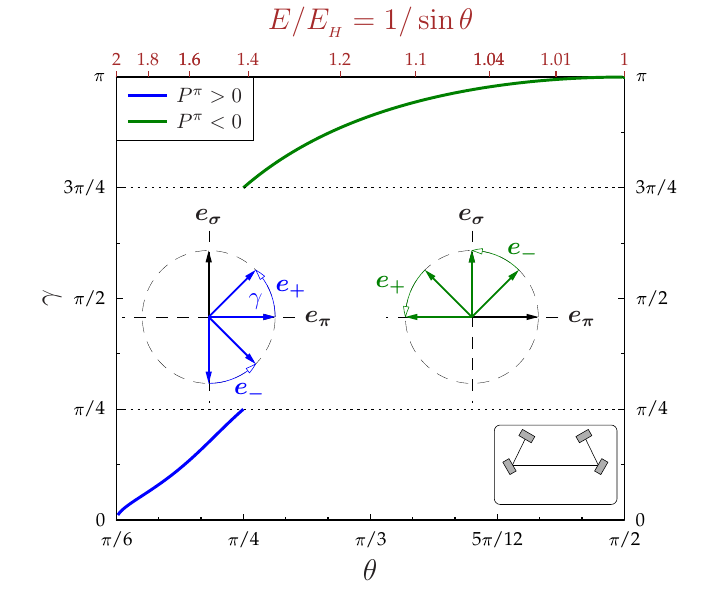}  
\caption{Rotation angle $\gamma$ of the mirror-symmetric-cavity
eigenpolarization basis $\{\epp,\epm\}$ relative to the intrinsic
basis $\{\esi,\epi\}$ as a function of Bragg angle $\theta$. The angle
$\gamma$ is defined by Eq.~\eqref{eq091} for $P^\pi>0$ and by
Eq.~\eqref{eq093} for $P^\pi<0$. In the twofold rotationally symmetric
cavity, the leading eigenpolarization basis coincides with
$\{\esi,\epi\}$ for $P^\pi>0$. For $P^\pi<0$, the zeroth-order
reflection matrix is diagonalized by the circular basis of
Eq.~\eqref{eq0A2170b}; finite-$\eta$ corrections generally modify
these states.
}
\label{fig8} 
\end{figure}

\subsubsection{Small-deviation interpretation}

The numerical results in Fig.~\ref{fig51} can be understood from the
small-$\eta$ structure of the cavity reflection matrices derived in
Appendix~\ref{smalleta}. For $P^\pi>0$, the cross-polarization
amplitude of the mirror-symmetric cavity is linear in $\eta$,
\[
R_{\ind{6m}}^{\pi\sigma}=O(\eta),
\]
whereas in the twofold rotationally symmetric cavity the
leading-order contribution cancels between the two backscattering
units,
\[
R_{\ind{6r}}^{\pi\sigma}=O(\eta^2).
\]
Consequently, the corresponding cross-polarization intensities scale
as
\[
|R_{\ind{6m}}^{\pi\sigma}|^2=O(\eta^2),
\qquad
|R_{\ind{6r}}^{\pi\sigma}|^2=O(\eta^4).
\]
This additional suppression explains the substantially weaker
polarization mixing in the rotationally symmetric configuration below
$\theta=45^\circ$.

For $P^\pi<0$, the dominant and subdominant co-polar amplitude combinations $u$
and $v$  exchange roles. In this range the mirror-symmetric
configuration again suppresses the cross-polarization amplitude at the
center of the common total-reflection region,
\[
R_{\ind{6m}}^{\pi\sigma}=O(\eta),
\]
whereas the rotationally symmetric configuration generally retains a
finite cross-polarization amplitude already at $\eta=0$,
\[
R_{\ind{6r}}^{\pi\sigma}=O(\sin12\varphi).
\]
Thus, the mirror-symmetric configuration provides the more favorable
polarization behavior above $\theta=45^\circ$.

This has a clear practical implication. The two cavity configurations
provide complementary polarization behavior. For $P^\pi>0$, the
twofold rotationally symmetric configuration suppresses
cross-polarization mixing more strongly near the center of the common
total-reflection region, because the leading cross-polarization
amplitude is canceled between the two backscattering units. For
$P^\pi<0$, the mirror-symmetric configuration is more favorable,
whereas the rotationally symmetric configuration generally retains
substantial polarization mixing even at the center of the reflection
region. These results identify the Bragg-angle ranges in which each
cavity symmetry provides the more favorable intrinsic-polarization
response.

\subsection{Eigenpolarizations, phases, and time delays}

\subsubsection{Eigenpolarizations}

The intrinsic polarization basis $\{\esi,\epi\}$, defined in
Eq.~\eqref{eq051}, is not in general an eigenpolarization basis of the
cavity. Exceptions are the coplanar limit at $\theta=\pi/6$, the
backscattering limit $\theta\to\pi/2$, and, for the
mirror-symmetric cavity, the center of the reflection region
$\eta=0$. Away from these limits the intrinsic polarizations are
coupled.

A central question is whether the cavity eigenpolarizations can be
chosen independent of $\eta$. In general, the exact eigenvectors of
the six-crystal reflection matrices depend on $\eta$. However, as
shown in Appendix~\ref{eigen}, if terms of order $\eta^3$ and higher
are neglected, for the mirror-symmetric cavity $\hat R_{\ind{6m}}$  can be
written in the form
\begin{equation}
\hat R_{\ind{6m}}
=
R_0 \hat I + \lambda \hat M + O(\eta^3),
\end{equation}
where $\hat I$ is the identity matrix, $R_0$ and $\lambda$ are scalar
functions of $\eta$, and $\hat M$ is independent of $\eta$. Therefore,
within the $O(\eta^2)$ approximation and away from degeneracies, the
eigenvectors of $\hat R_{\ind{6m}}$ coincide with those of $\hat M$
and are independent of $\eta$.

Therefore, for the mirror-symmetric cavity, the resulting
$\eta$-independent eigenpolarization basis
$\{\epp,\epm\}$ is a linear basis related to the intrinsic basis by
\begin{equation}
\begin{pmatrix}
\epp \\
\epm
\end{pmatrix}
=
\hat W(\gamma)
\begin{pmatrix}
\esi \\
\epi
\end{pmatrix},\!\!\!
\quad
\hat W(\gamma)
=
\begin{pmatrix}
\sin\gamma & \cos\gamma \\
-\cos\gamma & \sin\gamma
\end{pmatrix},
\label{eq0915b}
\end{equation}
where $\gamma$ is given by Eq.~\eqref{eq091} for $P^\pi>0$ and by
Eq.~\eqref{eq093} for $P^\pi<0$. Figure~\ref{fig8} shows the rotation
angle $\gamma$ as a function of $\theta$.

The apparent jump of $\gamma$ by $\pi/2$ at $\theta=\pi/4$ in
Fig.~\ref{fig8} reflects only an interchange of the two orthogonal
eigenpolarization branches in the mirror-symmetric case. Since
$P^\pi=\cos 2\theta$ changes sign at this point, the labeling of
$\epp$ and $\epm$ changes, while the unordered pair of physical axes
remains continuous.

For the twofold rotationally symmetric cavity, the structure of $\hat
R_{\ind{6r}}$ is
different. If $P^\pi>0$, Appendix~\ref{eigen} shows that
$\hat R_{\ind{6r}}$ is diagonal in the intrinsic basis through order
$\eta^2$, so the eigenpolarizations are simply $\epp=\esi$  and $\epm=\epi$. If
$P^\pi<0$, the zeroth-order part of $\hat R_{\ind{6r}}$ already has
finite off-diagonal terms, and
the zeroth-order eigenpolarizations are circular polarization states
\begin{equation}
\begin{pmatrix}
    \epp \\
    \epm 
\end{pmatrix}
=\hat{W}
\begin{pmatrix}
    \esi \\
    \epi
\end{pmatrix}
,
\quad
\hat{W}
= \frac{1}{\sqrt 2} 
\begin{pmatrix}
1 & -i \\
1 & i
\end{pmatrix}.
\label{eq0A2170b} 
\end{equation}
The
$\eta$-dependent correction, however, is not generally diagonal in
this circular basis. Thus, the circular states provide a zeroth-order
description of the cavity eigenpolarizations, while the exact
eigenpolarizations generally acquire $\eta$-dependent corrections.

Thus, the two cavity symmetries lead to qualitatively different
polarization structures. The eigenstates of $\hat R_{\ind{6m}}$ are
rotated linear combinations of $\esi$ and $\epi$ with a
$\theta$-dependent rotation angle. For $\hat R_{\ind{6r}}$, the
intrinsic linear polarizations provide the leading eigenbasis for
$P^\pi>0$, whereas circular polarizations diagonalize only the
zeroth-order matrix for $P^\pi<0$.

\begin{figure*}[t!]
  \includegraphics[width=1.0\textwidth]{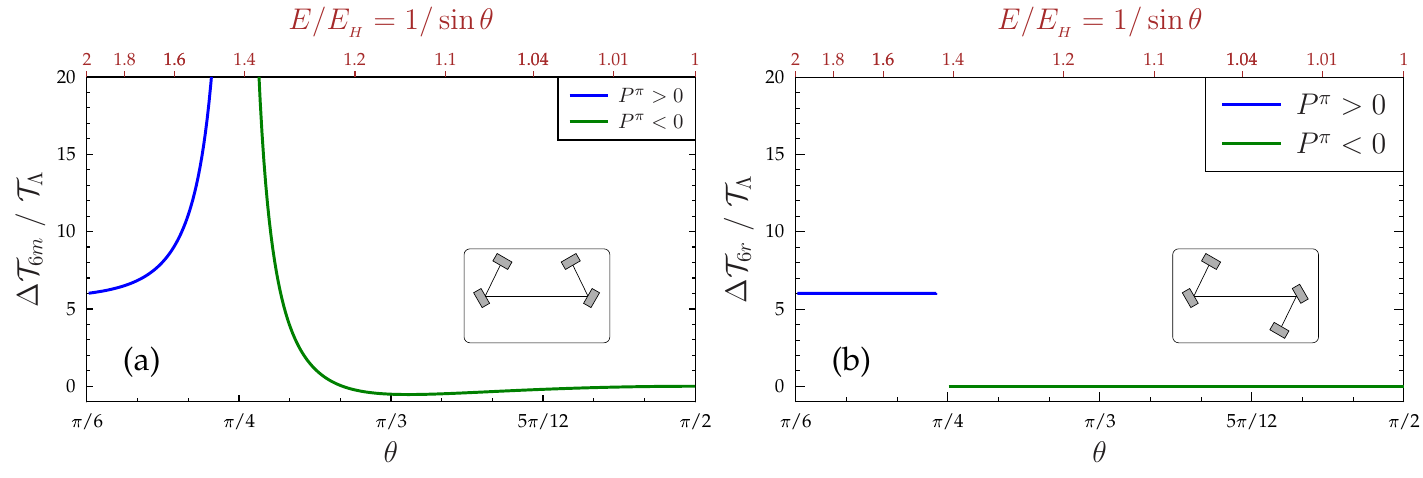}  
  \caption{Relative first-order time delay
    $\Delta{\mathcal T}_{\ind{6k}}$ ($k=m,r$) between the relevant
    cavity polarization modes after one round trip, in units of the
    single-crystal delay time $\timex$, Eq.~\eqref{eq00826}, as a
    function of Bragg angle $\theta$. (a) Mirror-symmetric cavity:
    $\Delta\timesix$ is defined in Eq.~\eqref{eq0126}. (b) Twofold
    rotationally symmetric cavity: $\Delta\timesixr$ is given by
    Eq.~\eqref{eq0126r}; for $P^\pi<0$, it refers to the zeroth-order
    circular modes.}
  \label{fig9} 
\end{figure*}

\begin{turnpage}
  \begin{figure*}[p]
    \centering
\includegraphics[width=1.3\textwidth]{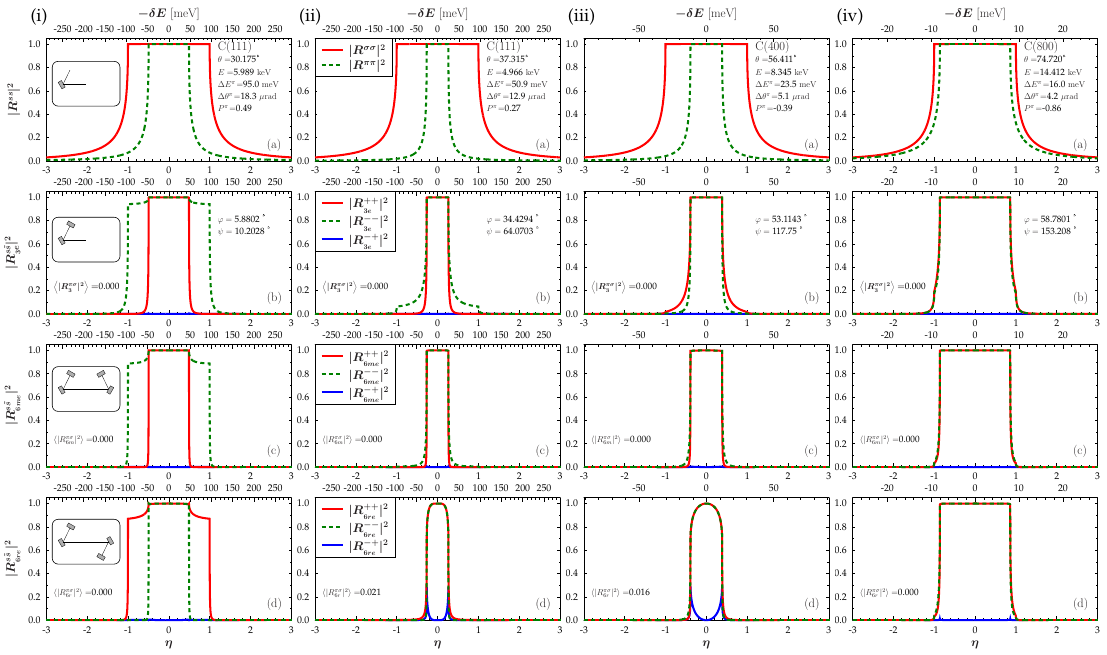}
\caption{Similar to Fig.~\ref{fig51}, but with the reflection matrices
  expressed in the corresponding eigenpolarization bases
  $\{\epp,\epm\}$ associated with the cavity modes.  For the
  mirror-symmetric cavity, this is the rotated linear basis of
  Eq.~\eqref{eq0915b}. For the twofold rotationally symmetric cavity, it is
  the intrinsic basis for $P^\pi>0$ and the zeroth-order circular
  basis of Eq.~\eqref{eq0A2170b} for $P^\pi<0$.}
\label{fig14}
\end{figure*}
\end{turnpage}

\subsubsection{Eigenphases and time delays}

The matrices $\hat R_{\ind{3}}$, $\hat R_{\ind{6m}}$, and
$\hat R_{\ind{6r}}$ are unitary in the common total-reflection region,
because they are products of unitary single-crystal reflection matrices
and unitary polarization-rotation matrices. Their eigenvalues
therefore have unit modulus and may be written as
\begin{equation}
\rho_\pm(\eta)=\exp\left[i\delta_\pm(\eta)\right],
\end{equation}
with real eigenphases $\delta_\pm(\eta)$. For the mirror-symmetric
cavity, and for the twofold rotationally symmetric cavity with
$P^\pi>0$, the small-$\eta$ expansion in Appendix~\ref{eigen}  gives
\begin{equation}
\delta_\pm(\eta)=\pi-A_\pm\eta+O(\eta^3).
\label{eq0A2041}
\end{equation}
In particular, for the mirror-symmetric cavity 
\begin{equation}
A_\pm=
\begin{cases}
-6\beta\pm 2\alpha q, & \text{if } P^\pi>0,\\
-6\beta\pm 2\alpha T, & \text{if } P^\pi<0.
\end{cases}
\label{eq0122}
\end{equation}
Using the time-energy representation of $\eta$,
Eqs.~\eqref{eq00824}--\eqref{eq00826}, the Bragg-diffraction delay of
the two eigenpolarizations is
\begin{equation}
{\mathcal T}_\pm=-A_\pm \timex
=
\timesix \mp \frac{\Delta \timesix}{2},
\label{eq0124}
\end{equation}
where $\timesix=6\beta\timex$. The relative delay is therefore
\begin{equation}
\Delta \timesix=
\begin{cases}
4|\alpha| q\,\timex, & \text{if } P^\pi>0,\\
4|\alpha| T\,\timex, & \text{if } P^\pi<0.
\end{cases}
\label{eq0126}
\end{equation}

For the twofold rotationally symmetric cavity
\[
  A_\pm= -6\beta\pm 3,  \text{~if } P^\pi>0.
\]
If $P^\pi<0$, the
zeroth-order circular modes have an additional phase offset
$\mp12\varphi$, while their common first-order phase correction is
$6\beta\eta$ resulting in
\[
\delta_\pm(\eta)= \pi\mp12\varphi+6\beta\eta+O(\eta^2).
\]
Finite-$\eta$ corrections modify the eigenvectors and may
contribute to the mode splitting at higher order.

The relative delay in the rotationally symmetric cases is therefore
\begin{equation}
\Delta \timesixr=
\begin{cases}
6\timex, & \text{if } P^\pi>0,\\
0+O(\eta), & \text{if } P^\pi<0.
\end{cases}
\label{eq0126r}
\end{equation}

Figure~\ref{fig9}(a) shows $\Delta\timesix$ for the mirror-symmetric
cavity. In that case the relative delay grows strongly as $\theta$
approaches $\pi/4$ and becomes small for $\theta>\pi/3$. In the
twofold rotationally symmetric cavity, the corresponding dependence
$\Delta\timesixr$ in Fig.~\ref{fig9}(b) is instead
$\theta$-independent but having different values in the $0<\theta<\pi/4$
($P^\pi>0$), and in the $\pi/4<\theta<\pi/2$ ($P^\pi<0$) ranges. These
delays must be taken into account when matching the cavity round-trip
time.

\subsubsection{Reflectivity in the eigenpolarization basis}

The matrices $\hat R_{\ind{3}}$, $\hat R_{\ind{6m}}$, and
$\hat R_{\ind{6r}}$, originally defined in the intrinsic basis
$\{\esi,\epi\}$, may be transformed to the corresponding cavity
eigenpolarization basis $\{\epp,\epm\}$ as
\begin{equation}
\hat R_{\ind{ke}}
=
\hat W \hat R_{\ind{k}} \hat W^{-1},
\qquad
k=3,6m,6r .
\end{equation}

For the mirror-symmetric cavity,
$\hat W=\hat W(\gamma)$ is given by Eq.~\eqref{eq0915b}, and
$\hat W^{-1}(\gamma)=-\hat W(-\gamma)$. The transformed elements
(with $k=3$ or $6m$) are
\begin{equation}
\begin{split}
R_{\ind{ke}}^{++}=& R_{\ind{k}}^{\sigma\sigma}\sin^{2}\gamma +
                    R_{\ind{k}}^{\pi\pi}\cos^{2}\gamma + R_{\ind{k}}^{\sigma\pi}\sin2\gamma,
\\
R_{\ind{ke}}^{+-}=& R_{\ind{ke}}^{-+}  = -R_{\ind{k}}^{\sigma\pi}\cos2\gamma
+ \frac{R_{\ind{k}}^{\pi\pi}-R_{\ind{k}}^{\sigma\sigma}}{2}\sin2\gamma,
\\
R_{\ind{ke}}^{--}=& R_{\ind{k}}^{\sigma\sigma}\cos^{2}\gamma +
                    R_{\ind{k}}^{\pi\pi}\sin^{2}\gamma -
R_{\ind{k}}^{\sigma\pi}\sin2\gamma.
\end{split}
\label{eq0150}
\end{equation}
For the twofold rotationally symmetric cavity with $P^\pi>0$, the
eigenbasis coincides with the intrinsic basis, so no transformation is
required. For $P^\pi<0$, the transformation is instead to the circular
basis in Eq.~\eqref{eq0A2170b} and the transformed elements are
\begin{equation}
\begin{split}
2R_{\ind{6re}}^{++}={}&
R_{\ind{6r}}^{\sigma\sigma}+R_{\ind{6r}}^{\pi\pi}
+i\left(
R_{\ind{6r}}^{\sigma\pi}-R_{\ind{6r}}^{\pi\sigma}
\right),\\
2R_{\ind{6re}}^{+-}={}&
R_{\ind{6r}}^{\sigma\sigma}-R_{\ind{6r}}^{\pi\pi}
-i\left(
R_{\ind{6r}}^{\sigma\pi}+R_{\ind{6r}}^{\pi\sigma}
\right),\\
2R_{\ind{6re}}^{-+}={}&
R_{\ind{6r}}^{\sigma\sigma}-R_{\ind{6r}}^{\pi\pi}
+i\left(
R_{\ind{6r}}^{\sigma\pi}+R_{\ind{6r}}^{\pi\sigma}
\right),\\
2R_{\ind{6re}}^{--}={}&
R_{\ind{6r}}^{\sigma\sigma}+R_{\ind{6r}}^{\pi\pi}
-i\left(
R_{\ind{6r}}^{\sigma\pi}-R_{\ind{6r}}^{\pi\sigma}
\right).
\end{split}
\label{eq01505}
\end{equation}

Figure~\ref{fig14} shows the squared moduli of the three-crystal and
six-crystal reflection-matrix elements in the corresponding cavity
eigenpolarization bases for the same Bragg angles as in
Fig.~\ref{fig51}. The transformation to these bases shows that the
strong polarization conversion observed in the intrinsic basis does
not necessarily constitute an intrinsic cavity loss mechanism.

For the mirror-symmetric cavity, the eigenpolarization basis strongly
suppresses cross-polarization mixing over a broad range of $\eta$, as
shown in columns (b) and (c), although the basis itself is obtained
from the small-$\eta$ approximation. The diagonal reflectivities
therefore remain close to unity throughout most of the common
total-reflection interval. If the incident radiation is prepared in a
cavity eigenpolarization state, the polarization-mixing constraint on
the usable tuning range can consequently be substantially relaxed. The
remaining limitation near $\theta=\pi/4$ arises primarily from the
collapse of the common total-reflection interval as $P^\pi\to0$.

The improvement is less pronounced for the twofold rotationally
symmetric cavity. In the range $\pi/6<\theta<\pi/4$ ($P^\pi>0$), the
cavity eigenpolarization basis coincides with the intrinsic basis and
therefore provides no additional reduction of polarization mixing. In
the range $\pi/4<\theta<\pi/2$ ($P^\pi<0$), transformation to the
zeroth-order circular basis likewise does not substantially reduce the
mixing. This behavior is consistent with the fact that the circular
states diagonalize only the zeroth-order reflection matrix, while the
finite-$\eta$ corrections generally modify the cavity
eigenpolarizations.

\section{Cavity tuning range}

\begin{figure}[t!]
  \includegraphics[width=0.50\textwidth]{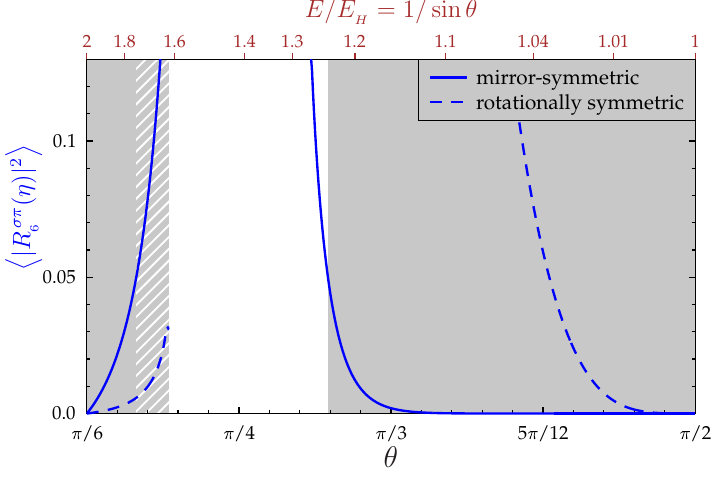}
  \caption{Average admixture
    $\langle |R^{\sigma\pi}_6(\eta)|^2 \rangle$
    of the complementary polarization component after one round trip,
    evaluated over $|\eta|\leq 1/4$, as a function of Bragg angle
    $\theta$ and relative photon energy $E/E_{\ind{H}}$. Solid lines
    show numerical results for the mirror-symmetric cavity based on
    Eq.~\eqref{eq054}, and dashed lines those for the twofold rotationally
    symmetric cavity based on Eq.~\eqref{eq054r}. The white interval,
    from $E_{\ind{1}}=1.239E_{\ind{H}}$
    ($\theta_{\ind{1}}=53.79^\circ$) to
    $E_{\ind{2}}=1.622E_{\ind{H}}$ ($\theta_{\ind{2}}=38.07^\circ$),
    is excluded because the polarization mixing exceeds 5\%. This
    reduces the full tuning range, from $E=E_{\ind{H}}$ at
    $\theta=90^\circ$ to $E=2E_{\ind{H}}$ at $\theta=30^\circ$, to
    61.7\%. Solid shading marks the usable range for the
    mirror-symmetric cavity; hatched shading marks the additional
    extension provided by the twofold rotationally symmetric cavity.}
  \label{fig5}
\end{figure}

\begin{figure*}[t!]
  \includegraphics[width=0.99\textwidth]{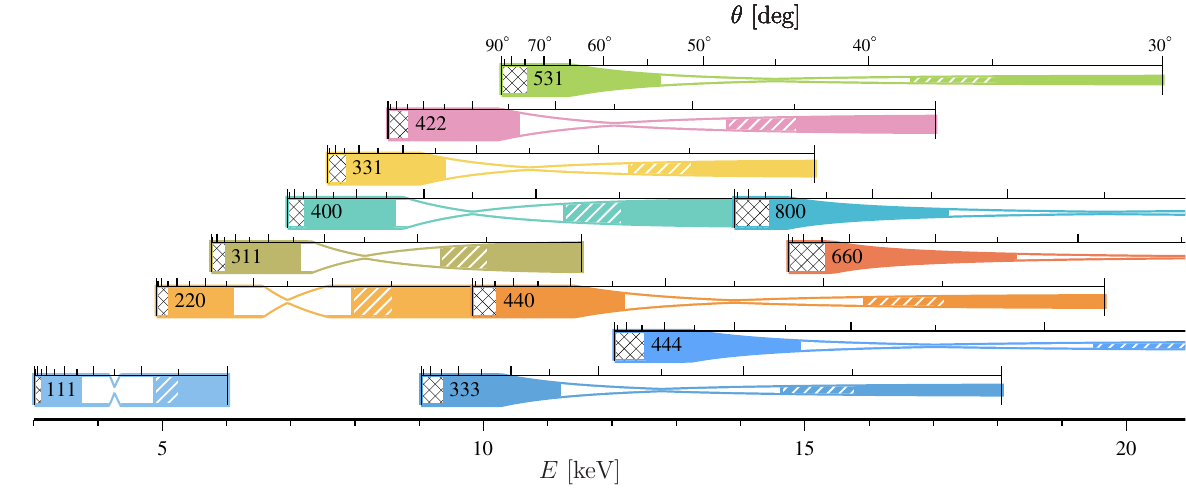}
  \caption{Photon-energy tunability of the six-crystal noncoplanar
    cavity for different diamond Bragg reflections $hkl$. For each
    reflection, the tuning range spans from $E_{\ind{H}}$ at
    $\theta=90^\circ$ to $2E_{\ind{H}}$ at $\theta=30^\circ$, where
    $\vc H=(h,k,l)$ and the values of $E_{\ind{H}}$ are taken from
    Ref.~\cite{SL12}. The cross-hatched 3.5\% portion is excluded by
    geometrical constraints. The usable fraction of the tuning range,
    approximately 58\%, for which cross-polarization mixing in the
    intrinsic polarization basis remains below 5\%, is indicated by
    color shading for each fundamental and harmonic reflection: solid
    for the mirror-symmetric cavity and hatched for the twofold rotationally
    symmetric cavity. Regions where the polarization mixing exceeds
    5\% are shown in white. Where the common total-reflection width
    $\Delta\theta^\pi$, Eq.~\eqref{eq-epspi}, falls below 3~$\mu$rad,
    the width of the corresponding region is reduced proportionally,
    indicating potential limitations. In the cavity eigenpolarization
    basis, cross-polarization mixing can be suppressed, leaving the
    usable fraction limited only by the reduced angular acceptance.}
  \label{fig6}
\end{figure*}

The photon-energy tuning range of a six-bounce noncoplanar cavity
using the same Bragg reflection in all six crystals can, in principle,
reach 100\% by varying the Bragg angle $\theta$. It extends from the
backreflection energy $E=E_{\ind{H}}$ at $\theta=90^\circ$ to
$E=2E_{\ind{H}}$ at $\theta=30^\circ$, as illustrated in
Fig.~\ref{fig3}. In practice, this range is limited by several
constraints.

\subsubsection{Limitations}

\paragraph{Geometrical constraint.}
As discussed in Sec.~\ref{kinematics} and illustrated in
Fig.~\ref{fig7}, the longitudinal dimensions of the backscattering
units and the required longitudinal translations become excessively
large as $\theta\to 90^\circ$. It is therefore preferable to omit the
range $90^\circ \geq \theta \gtrsim 75^\circ$, hatched in
Fig.~\ref{fig7}. Because the photon energy varies over this range only
from $E_{\ind{H}}$ to approximately $1.035E_{\ind{H}}$, its exclusion
reduces the tuning range by only about 3.5\%.

\paragraph{Polarization mixing.}
If the undulator generates x-rays in one of the intrinsic linear
polarization states $\esi$ or $\epi$, defined in
Eq.~\eqref{eq051}, polarization mixing transfers part of the
radiation into the complementary state. This reduces the intensity in
the desired polarization and acts as an additional cavity loss.

Figure~\ref{fig5} shows the average round-trip admixture
$\langle |R^{\sigma\pi}_6(\eta)|^2 \rangle =
\langle |R^{\pi\sigma}_6(\eta)|^2 \rangle$, evaluated over
$|\eta|\leq 1/4$, as a function of $\theta$ and
$E/E_{\ind{H}}$. This quantity measures the cavity loss due to
polarization coupling. In the following, a 5\% admixture is adopted
as the limit for estimating the allowable range of $\theta$. As
$\theta$ approaches $45^\circ$, the polarization coupling becomes
significant while the common Bragg-reflection region simultaneously
narrows. Solid lines show numerical results for the mirror-symmetric
cavity based on Eq.~\eqref{eq054}, whereas dashed lines show the
corresponding results for the twofold rotationally symmetric cavity based on
Eq.~\eqref{eq054r}.

In the $P^\pi>0$ range, polarization mixing is substantially weaker
for the twofold rotationally symmetric cavity. The dashed curve terminates at
$\theta\simeq 38^\circ$, where the common total-reflection range
becomes narrower than the averaging interval. Consequently, the
averaged admixture decreases before reaching the 5\% limit.

In the $P^\pi<0$ range, polarization mixing is substantially weaker
for the mirror-symmetric cavity. The 5\% limit is reached at
$\theta \simeq 53.79^\circ$.

Excluding photon energies with polarization mixing above 5\% reduces
the tuning range to approximately 61.7\%. The excluded interval
extends from $E_{\ind{1}}=1.239E_{\ind{H}}$
($\theta_{\ind{1}}=53.79^\circ$) to
$E_{\ind{2}}=1.622E_{\ind{H}}$
($\theta_{\ind{2}}=38.07^\circ$). Together, the geometrical and
polarization-mixing constraints reduce the usable tuning range from
100\% to approximately 58\%.

\paragraph{Angular acceptance.}
So far, we have assumed a negligible angular spread of the x-rays in
the cavity. If this spread becomes comparable to or larger than the
common angular range of total Bragg reflection,
\begin{equation}
\Delta\theta^\pi\simeq
\epsilon_{\ind{H}}\tan\theta\,|\cos 2\theta|,
\label{eq-epspi}
\end{equation}
the tuning range is further restricted. Near $\theta=45^\circ$, where
polarization mixing is also substantial, this limitation is universal
and independent of the Bragg reflection. Elsewhere, as illustrated in
Fig.~\ref{fig6}, its impact is broader for high-index reflections,
for which the relative spectral-width invariant
$\epsilon_{\ind{H}}$ is small.

\subsubsection{Extensions}

Several approaches can extend the accessible photon-energy range.

\paragraph{Fundamental and harmonic reflections.}
Reflections having the same plane normal, such as diamond 111, 333,
and 444 reflections, can be used as fundamental and harmonic
reflections. If the angular-acceptance limitation is disregarded, this
extends the accessible range from approximately 3~keV to 18~keV.
Although the resulting range is not continuous, it covers about
7.8~keV in total, corresponding to an effective tuning range of about
250\%, as illustrated in Fig.~\ref{fig6}. Most gaps could be closed by
mounting crystals with different cuts, such as (220), (311), and
(400), on the same cavity crystal holders.

\paragraph{Operation in the eigenpolarization basis.}
The intervals excluded by high cross-polarization mixing, which account
for about 42\% of the tuning range for each reflection, can be
reduced if the undulator generates x-rays in one of the
eigenpolarization states of the mirror-symmetric cavity, $\epp$ or
$\epm$. Polarization mixing is then suppressed, and the usable range is
limited primarily by the reduced angular acceptance. This can be
achieved by: (i) using an APPLE-II or similar undulator with variable
linear polarization~\cite{HWANG99,YAM02,QHD24}; (ii) using
transmission phase retarders, such as diamond phase
plates~\cite{GMG94,SIY14}; or (iii) rotating the principal cavity
plane $(x,z)$ about the $\vc{z}$ axis to align the polarization from a
conventional planar undulator with a cavity eigenpolarization.

In the second approach, a phase retarder placed downstream of a planar
undulator rotates the x-ray linear polarization by the required angle
$\gamma$, while a complementary retarder placed upstream restores the
initial polarization. Transmission phase plates are not lossless,
however, owing to residual Bragg diffraction, diamond absorption, and
other imperfections. In the third approach, the required rotation
angle may be written as $\gamma$ or as an equivalent angle
corresponding to the same eigenpolarization axis, depending on the
branch convention; see Fig.~\ref{fig8}.

In principle, these approaches make the intervals excluded by high
cross-polarization mixing accessible. In practice, the usable range
remains limited near $\theta=45^\circ$, because the common
total-reflection interval, $|\eta|\leq |P^\pi|$, collapses to
$\eta=0$ as $P^\pi\to 0$. The corresponding energy and angular
acceptances then become too narrow.

\begin{table*}
\begin{tabular}{|llllllllll|}
\hline
  $hkl$ & $E$ & $\theta$  & $\Delta E^{\pi}$ & $\Delta\theta^{\pi}$ & $P^{\pi}$ &  $\langle|R^{\pi\sigma}_{\ind{3}}|^2\rangle$ &  $\langle|R^{\pi\sigma}_{\ind{6m}}|^2\rangle$ &  $\langle|R^{\pi\sigma}_{\ind{6r}}|^2\rangle$  & edge/ \\
 & [keV] & [deg]  & [meV] & [$\mu$rad] &  & &   &  & resonance  \\  
  \hline
111 &      3.179   &   71.253   &    152.3   &    149.1   &    -0.79   &   0.0380   &   0.0000   &   0.1462   &  Pd~$L_3$  \\
111 &      3.205   &   69.928   &    146.8   &    133.4   &    -0.76   &   0.0506   &   0.0000   &   0.1921   &  Ar~$K$    \\
111 &      3.608   &   56.548   &     75.3   &     37.9   &    -0.39   &   0.4282   &   0.0116   &   0.9678   &  K~$K$     \\
111 &      4.058   &   47.887   &     19.3   &      7.1   &    -0.10   &   0.3725   &   0.1176   &   0.0936   &  Ca~$K$    \\
111 &      4.492   &   42.079   &     19.5   &      5.9   &     0.10   &   0.0613   &   0.1262   &   0.0044   &  Sc~$K$    \\
111 &      4.966   &   37.315   &     50.9   &     12.9   &     0.27   &   0.0418   &   0.1331   &   0.0205   &  Ti~$K$    \\
111 &      5.465   &   33.425   &     75.5   &     16.5   &     0.39   &   0.0075   &   0.0266   &   0.0029   &  V~$K$     \\
111 &      5.989   &   30.175   &     95.0   &     18.3   &     0.49   &   0.0002   &   0.0008   &   0.0001   &  Cr~$K$    \\
333 &      9.569   &   70.684   &     21.3   &      6.7   &    -0.78   &   0.0431   &   0.0000   &   0.1648   &  Zn~$K$    \\
444 &     12.389   &   76.376   &     22.2   &      7.6   &    -0.89   &   0.0101   &   0.0000   &   0.0401   &  $^{45}$Sc \\
220 &      4.966   &   81.831   &    101.7   &    144.2   &    -0.96   &   0.0013   &   0.0000   &   0.0051   &  Ti~$K$    \\
220 &      5.465   &   64.089   &     65.5   &     27.4   &    -0.62   &   0.1486   &   0.0002   &   0.5059   &  V~$K$     \\
220 &      5.989   &   55.162   &     36.8   &     10.8   &    -0.35   &   0.4999   &   0.0239   &   0.9761   &  Cr~$K$    \\
220 &      8.345   &   36.090   &     32.4   &      4.8   &     0.31   &   0.0240   &   0.0814   &   0.0103   &  Ni~$K$    \\
220 &      8.944   &   33.339   &     42.0   &      5.6   &     0.40   &   0.0072   &   0.0256   &   0.0027   &  Yb~$L_3$  \\
220 &      8.990   &   33.147   &     42.6   &      5.7   &     0.40   &   0.0065   &   0.0233   &   0.0025   &  Cu~$K$    \\
220 &      9.569   &   30.911   &     50.1   &      6.1   &     0.47   &   0.0013   &   0.0047   &   0.0005   &  Zn~$K$    \\
440 &     10.367   &   71.497   &     28.7   &      8.7   &    -0.80   &   0.0360   &   0.0000   &   0.1387   &  Ga~$K$    \\
400 &      6.977   &   85.120   &     59.7   &    100.6   &    -0.99   &   0.0002   &   0.0000   &   0.0006   &  Eu~$L_3$  \\
400 &      7.130   &   77.156   &     54.6   &     34.5   &    -0.90   &   0.0080   &   0.0000   &   0.0316   &  Fe~$K$    \\
400 &      7.720   &   64.220   &     37.7   &     11.2   &    -0.62   &   0.1455   &   0.0002   &   0.4970   &  Co~$K$    \\
400 &      8.345   &   56.411   &     23.5   &      5.1   &    -0.39   &   0.4351   &   0.0125   &   0.9707   &  Ni~$K$    \\
400 &      8.944   &   51.008   &     12.6   &      2.2   &    -0.21   &   0.6206   &   0.1483   &   0.5534   &  Yb~$L_3$  \\
400 &      8.990   &   50.648   &     11.9   &      2.1   &    -0.20   &   0.6032   &   0.1496   &   0.4855   &  Cu~$K$    \\
800 &     14.413   &   74.720   &     16.0   &      4.2   &    -0.86   &   0.0163   &   0.0000   &   0.0640   &  $^{57}$Fe \\
\hline    
\end{tabular}
\caption{Representative operating points for the six-bounce
noncoplanar cavity using fundamental and harmonic diamond reflections
from the 111, 220, and 400 families. Here $E$ is the central photon
energy, $\theta$ is the corresponding Bragg angle, and
$\Delta E^\pi$ and $\Delta\theta^\pi$ are the spectral width and
angular acceptance, respectively, for $\pi$-polarized x-rays (and of the cavity as a whole). The
quantity $P^\pi$ is the $\pi$-polarization factor, and
$\langle|R_{\ind{k}}^{\pi\sigma}|^2\rangle$ is the polarization-mixing
factor for $k=3,6m,6r$, averaged over $|\eta|\leq1/4$. The final column
lists selected atomic absorption edges and nuclear resonances near the
corresponding photon energies.}
\label{tab1}
\end{table*}

Table~\ref{tab1} provides, for reference, examples of photon energies
accessible with the six-crystal noncoplanar cavity using fundamental
and harmonic diamond reflections from the 111, 220, and 400 families,
including selected atomic absorption edges and nuclear resonances.

\section{Conclusions}

We have studied a compact, tunable, noncoplanar x-ray cavity for
cavity-based x-ray free-electron lasers. The cavity consists of six
Bragg-reflecting crystals arranged as two three-crystal
backscattering units in either a mirror-symmetric or a twofold
rotationally symmetric configuration. Compared with planar bowtie
cavities, the three-dimensional geometry substantially broadens the
photon-energy tuning range while retaining a compact transverse
footprint suitable for installation in accelerator tunnels.

We derived analytical expressions for the cavity geometry, crystal
kinematics, photon-energy tuning, and polarization-dependent
Bragg-reflection matrices. For a given Bragg reflection, the ideal
tuning range extends from $E=E_{\ind{H}}$ at backscattering
($\theta=90^\circ$) to $E=2E_{\ind{H}}$ at
$\theta=30^\circ$, corresponding to a 100\% photon-energy range.
In practice, geometrical constraints near backscattering and
polarization mixing near $\theta=45^\circ$ reduce the directly usable
range to approximately 58\%.

The intrinsic polarization basis defined by the cavity geometry is not
generally the eigenpolarization basis of the full six-crystal cavity.
Consequently, radiation initially prepared in one intrinsic linear
polarization state can acquire a complementary-polarization component,
resulting in an effective cavity loss. The two cavity symmetries
provide complementary intrinsic-polarization behavior: for $P^\pi>0$,
the twofold rotationally symmetric configuration more strongly
suppresses cross-polarization mixing near the center of the common
total-reflection region, whereas for $P^\pi<0$ the mirror-symmetric
configuration is more favorable. In the latter range, the rotationally
symmetric cavity generally retains substantial polarization mixing
even at the center of the reflection region.

In the small-deviation approximation, the mirror-symmetric cavity has
fixed linear eigenpolarizations, given by $\theta$-dependent rotations
of the intrinsic basis vectors $\esi$ and $\epi$. Its round-trip
eigenvalues have unit modulus in the common total-reflection region and
therefore differ primarily by $\eta$-dependent phases. Preparing the
incident radiation in a mirror-symmetric-cavity eigenpolarization
strongly suppresses cross-polarization mixing over a broad range of
$\eta$ and can substantially relax the associated tuning constraint.
The usable range then remains limited primarily by the rapidly reduced
angular acceptance near $\theta=45^\circ$, where $P^\pi\to0$ and the
common total-reflection interval collapses.

For the twofold rotationally symmetric cavity, the eigenpolarizations
coincide, to leading order, with the intrinsic linear polarizations in
the $P^\pi>0$ range and therefore provide no additional suppression of
polarization mixing. In the $P^\pi<0$ range, the zeroth-order matrix is
diagonalized by circular polarizations; finite-$\eta$ corrections,
however, generally modify these states, and the circular basis does not
substantially reduce the mixing. Thus, among the two configurations,
the mirror-symmetric cavity provides the more robust route to
operation in a fixed polarization eigenbasis.

Finally, fundamental and harmonic diamond reflections, together with
reflections from different crystallographic families, can substantially
extend the accessible photon-energy range. For example, combining 111,
333, and 444 reflections provides an effective tuning range of about
250\%, covering approximately 3--18~keV. Remaining gaps can be
reduced or closed by employing additional crystal cuts.

All quoted usable tuning ranges represent upper bounds and may be
further restricted when the angular acceptance of the common
total-reflection region of the Bragg reflections becomes small.

These results establish noncoplanar multi-crystal cavities as a
potentially viable route to compact, broadly tunable x-ray resonators
for future CBXFELs. The analytical framework developed here provides a
basis for optimizing cavity geometry, selecting Bragg reflections, and
managing polarization effects in practical implementations.

\section{Acknowledgments}
Work was supported by the U.S. Department of Energy, Office of
Science, Office of Basic Energy Sciences, under Contract
No. DE-AC02-06CH11357.


\appendix

\section{Polarization basis vectors}
\label{polarizationbasis}
  
First, using the Eqs.~\eqref{eq0100}-\eqref{eq0130} we calculate the wave vectors
\begin{equation}
\begin{split}
\vc{\kappa}_0=&\kappa(0,0,1),\\
\vc{\kappa}_1=&\kappa
\left(-\sin 2\theta\cos\varphi,\, \sin 2\theta\sin\varphi,\,\cos 2\theta\right),\\
\vc{\kappa}_2=&\kappa
\left(-\sin 2\theta\cos\varphi,\,-\sin 2\theta\sin\varphi,\,-\cos 2\theta\right),\\
\vc{\kappa}_3=&\kappa(0,0,-1),\\
\vc{\kappa}_4=&\kappa
\left(\sin 2\theta\cos\varphi,\, \pm \sin 2\theta\sin\varphi,\,-\cos 2\theta\right),\\
\vc{\kappa}_5=&\kappa
\left(\sin 2\theta\cos\varphi,\,\mp\sin 2\theta\sin\varphi,\,\cos 2\theta\right),\\
\vc{\kappa}_6=&\kappa(0,0,1).
\end{split}
\label{eq0A100}
\end{equation}
Therefore, $\vc{\kappa}_3=-\vc{\kappa}_0$ and $\vc{\kappa}_6=\vc{\kappa}_0$, as should be.

Using these results together with the definitions \eqref{eq0060} we
calculate the polarization basis for the first crystal,
\begin{equation}
\begin{split}
\vc{\sigma}_{10}
=&
\vc{\sigma}_{11}
=
(\sin\varphi,\cos\varphi,0),\\
\vc{\pi}_{10}
=&
(-\cos\varphi,\sin\varphi,0),\\
\vc{\pi}_{11}
=&
(-\cos 2\theta\cos\varphi,\,
\cos 2\theta\sin\varphi,\,
-\sin 2\theta);
\end{split}
\end{equation}
for the second crystal
\begin{equation}
\begin{split}
\vc{\sigma}_{20}
=&
\vc{\sigma}_{21}
=
(0,\cos\psi,-\sin\psi),\\
\vc{\pi}_{20}
=&
(-\sin\theta,\,
-\cos\theta\sin\psi,\,
-\cos\theta\cos\psi),\\
\vc{\pi}_{21}
=&
(\sin\theta,\,
-\cos\theta\sin\psi,\,
-\cos\theta\cos\psi);
\end{split}
\end{equation}
and for the third crystal
\begin{equation}
\begin{split}
\vc{\sigma}_{30}
=&
\vc{\sigma}_{31}
=
(-\sin\varphi,\cos\varphi,0),\\
\vc{\pi}_{30}
=&
(\cos 2\theta\cos\varphi,\,
\cos 2\theta\sin\varphi,\,
-\sin 2\theta),\\
\vc{\pi}_{31}
=&
(\cos\varphi,\sin\varphi,0).
\end{split}
\end{equation}

The intrinsic polarization basis is
\[
\esi=\hat{\vc{y}}=(0,1,0),
\qquad
\epi=-\hat{\vc{x}}=(-1,0,0).
\]
Thus, for an arbitrary vector
\[
\vc{v}=(v_x,v_y,v_z),
\]
one may write
\[
\vc{v}=v_y\esi-v_x\epi +v_z\hat{\vc{z}}.
\]
Therefore, for crystal C$_1$
\begin{equation}
\begin{split}
\vc{\sigma}_{10}
=&
\vc{\sigma}_{11}
=
\cos\varphi\,\esi
-
\sin\varphi\,\epi,\\
\vc{\pi}_{10}
=&
\sin\varphi\,\esi
+
\cos\varphi\,\epi,\\
\vc{\pi}_{11}
=&
\cos 2\theta\sin\varphi\,\esi
+
\cos 2\theta\cos\varphi\,\epi
-
\sin 2\theta\,\hat{\vc{z}}.
\end{split}
\end{equation}
Equivalently, by calculating the scalar products
$\vc{\sigma}_{10}\cdot \esi$, etc.,
\begin{equation}
\begin{pmatrix}
\vc{\sigma}_{10} \\
\vc{\pi}_{10}
\end{pmatrix}
=
\begin{pmatrix}
\cos\varphi & -\sin\varphi \\
\sin\varphi & \cos\varphi
\end{pmatrix}
\begin{pmatrix}
\esi \\
\epi
\end{pmatrix},
\label{eq0A2020}
\end{equation}
and
\begin{equation}
\begin{pmatrix}
\vc{\sigma}_{11} \\
\vc{\pi}_{11}
\end{pmatrix}
=
\begin{pmatrix}
\cos\varphi & -\sin\varphi \\
\cos 2\theta\sin\varphi & \cos 2\theta\cos\varphi
\end{pmatrix}
\begin{pmatrix}
\esi \\
\epi
\end{pmatrix},
\end{equation}

For crystal C$_2$
\begin{equation}
\begin{split}
\vc{\sigma}_{20}
=&
\vc{\sigma}_{21}
=
\cos\psi\,\esi
-
\sin\psi\,\hat{\mathbf z},\\
\vc{\pi}_{20}
=&
-\cos\theta\sin\psi\,\esi
+
\sin\theta\,\epi
-
\cos\theta\cos\psi\,\hat{\mathbf z},\\
\vc{\pi}_{21}
=&
-\cos\theta\sin\psi\,\esi
-
\sin\theta\,\epi
-
\cos\theta\cos\psi\,\hat{\mathbf z}.
\end{split}
\end{equation}
Equivalently,
\begin{equation}
\begin{pmatrix}
\vc{\sigma}_{20} \\
\vc{\pi}_{20}
\end{pmatrix}
=
\begin{pmatrix}
\cos\psi & 0 \\
-\cos\theta\sin\psi & \sin\theta
\end{pmatrix}
\begin{pmatrix}
\esi \\
\epi
\end{pmatrix},
\end{equation}
and
\begin{equation}
\begin{pmatrix}
\vc{\sigma}_{21} \\
\vc{\pi}_{21}
\end{pmatrix}
=
\begin{pmatrix}
\cos\psi & 0 \\
-\cos\theta\sin\psi & -\sin\theta
\end{pmatrix}
\begin{pmatrix}
\esi \\
\epi
\end{pmatrix},
\end{equation}

For crystal C$_3$,
\begin{equation}
\begin{split}
\vc{\sigma}_{30}
=&
\vc{\sigma}_{31}
=
\cos\varphi\,\esi
+
\sin\varphi\,\epi,\\
\vc{\pi}_{30}
=&
\cos 2\theta\sin\varphi\,\esi
-
\cos 2\theta\cos\varphi\,\epi
-
\sin 2\theta\,\hat{\mathbf z},\\
\vc{\pi}_{31}
=&
\sin\varphi\,\esi
-
\cos\varphi\,\epi.
\end{split}
\end{equation}
Equivalently,
\begin{equation}
\begin{pmatrix}
\vc{\sigma}_{30} \\
\vc{\pi}_{30}
\end{pmatrix}
=
\begin{pmatrix}
\cos\varphi & \sin\varphi \\
\cos 2\theta\sin\varphi & -\cos 2\theta\cos\varphi
\end{pmatrix}
\begin{pmatrix}
\esi \\
\epi
\end{pmatrix},
\end{equation}
and
\begin{equation}
\begin{pmatrix}
\vc{\sigma}_{31} \\
\vc{\pi}_{31}
\end{pmatrix}
=
\begin{pmatrix}
\cos\varphi & \sin\varphi \\
\sin\varphi & -\cos\varphi
\end{pmatrix}
\begin{pmatrix}
\esi \\
\epi
\end{pmatrix}.
\end{equation}

To calculate the relationship between basis vectors
$\{\vc{\sigma}_{\ind{20}},\vc{\pi}_{\ind{20}}\}$ and
$\{\vc{\sigma}_{\ind{11}},\vc{\pi}_{\ind{11}}\}$ we note that both
are transverse to the same wave vector $\vc{\kappa}_1$. Hence,
they are related by a two-dimensional rotation in the plane
perpendicular to $\vc{\kappa}_1$.  Taking scalar products
gives
\begin{equation}
\begin{split}
\vc{\sigma}_{20}\cdot\vc{\sigma}_{11}
=&
\frac{\cos 2\theta}{2\sin^2\theta},\qquad
\vc{\sigma}_{20}\cdot\vc{\pi}_{11}
=
\frac{\sin\varphi}{\sin\theta},\\
\vc{\pi}_{20}\cdot\vc{\sigma}_{11}
=&
-\frac{\sin\varphi}{\sin\theta}, \qquad
\vc{\pi}_{20}\cdot\vc{\pi}_{11}
=
\frac{\cos 2\theta}{2\sin^2\theta}.
\end{split}
\end{equation}
Because $\cos 2\theta/2\sin^2\theta=\cos 2\varphi$ and
$\sin\varphi/\sin\theta =\sin 2\varphi$, we write
\begin{equation}
  \begin{pmatrix}
    \vc{\sigma}_{20} \\
    \vc{\pi}_{20}
  \end{pmatrix}
  =\begin{pmatrix} \cos 2\varphi & \sin 2\varphi \\
    -\sin 2\varphi & \cos2\varphi \end{pmatrix}
  \begin{pmatrix}
    \vc{\sigma}_{11} \\
    \vc{\pi}_{11}
  \end{pmatrix}.
\label{eq0A2010}  
\end{equation}

\section{Small-$\eta$ approximations for Bragg-reflection polarization matrix}
\label{smalleta}

We derive the small-$\eta$ approximations for the elements of the
three-crystal, and six-crystal Bragg-reflection polarization matrices,
$R_{\ind{3}}^{s\tilde{s}}$, $R_{\ind{6m}}^{s\tilde{s}}$ and $R_{\ind{6r}}^{s\tilde{s}}$ in the
intrinsic polarization basis $\{\esi,\epi\}$, using the
second-order Taylor expansions in $\eta$ of the single-reflection
co-polar amplitude combinations $u$ and $v$, Eq.~\eqref{eq043}. These
second-order terms are required in order to preserve the normalization
identities
\[
\sum_{\tilde{s}=\sigma,\pi}|R_{\ind{k}}^{s\tilde{s}}|^2=1, \qquad
s=\sigma,\pi, \qquad k=3,6m,6r
\]
in the region of total reflection through order $\eta^2$.

In the following we use the abbreviations
\begin{equation}
C_n=\cos n\varphi,\quad
S_n=\sin n\varphi,\quad
T=1+2\cos 4\varphi ,
\end{equation}
\begin{equation}
\alpha=\frac{1}{2}\left(1-\frac{1}{|P^\pi|}\right),
\qquad
\beta=\frac{1}{2}\left(1+\frac{1}{|P^\pi|}\right).
\end{equation}

Using Eq.~\eqref{eq0020}, the
parameters $\alpha$ and $\beta$ can be written explicitly in terms of
$\varphi$ as
\begin{equation}
(\alpha,\beta)=
\begin{cases}
\left(-\dfrac{1}{2C_2},\,1+\dfrac{1}{2C_2}\right),
& 0<\theta<\pi/4,\\[10pt]
\left(1+\dfrac{1}{2C_2},\,-\dfrac{1}{2C_2}\right),
& \pi/4<\theta<\pi/2.
\end{cases}
\label{eqABphi}
\end{equation}

\subsection{Polarization factor $P^\pi>0$}
First, we consider the case $P^\pi>0$, corresponding to Bragg angles
in the range $\pi/6 \leq \theta<\pi/4$, for which the small-$\eta$
approximations for the co-polar amplitude combinations $u$ and $v$,
Eq.~\eqref{eq043} are
\begin{equation}
\begin{split}
u =& \alpha\eta +i\alpha\beta\,\eta^2 +O(\eta^3),\\
v=&-i+\beta\eta+\frac{i}{2}\left(\alpha^2+\beta^2\right)\eta^2+O(\eta^3),
\label{eq0B010}
\end{split}
\end{equation}
and therefore
\begin{equation}
  |u|\ll |v|.
\label{eq0B030}  
\end{equation}

Substituting these expressions into the definitions of $a$, $b$, and $d$ in Eq.~\eqref{eq052} and retaining terms through order $\eta^2$, we obtain
\begin{equation}
\begin{split}
a=&R_{\ind{3}}^{\sigma\sigma}= i-3\left(\beta-\frac{1}{2}\right)\eta -\\
& i\left[\left(\frac{5}{2}+2 C_4\right)\alpha^2+\frac{9}{2}\beta (\beta-1) \right]\eta^2+O(\eta^3),\\
b=&R_{\ind{3}}^{\pi\sigma}=R_{\ind{3}}^{\sigma\pi}= \alpha S_2\,\eta+3i\alpha\beta S_2\,\eta^2+O(\eta^3),\\
d=&R_{\ind{3}}^{\pi\pi}= i-3\left(\beta+\frac{1}{2}\right)\eta -\\
  &i\left[\left(\frac{5}{2}+2 C_4\right)\alpha^2+\frac{9}{2}\beta(\beta+1)\right]\eta^2+O(\eta^3).
\label{eq0B040}
\end{split}
\end{equation}

The squared moduli through order $\eta^2$ are then
\begin{equation}
\begin{split}
|a|^2=|d|^2 &= 1-\alpha^2 S_2^2\eta^2+O(\eta^3),\\
|b|^2 &=\alpha^2 S_2^2\eta^2+O(\eta^3). 
\label{eq0B050}
\end{split}
\end{equation}
Therefore, the squared matrix $\hat{R}_{\ind{3}}$ is normalized through order $\eta^2$.

In the small-deviation limit, $\eta \to 0$, the cross-polarization amplitude $|b|=|R_{\ind{3}}^{\pi\sigma}|$ vanishes, since $u \to 0$, and the three-crystal reflection matrix $\hat{R}_{\ind{3}}$ becomes diagonal.

Next, we use these results to calculate the small-$\eta$
approximations through order $\eta^2$ for the elements of the
six-crystal $R_{\ind{6m}}^{s\tilde{s}}$ and
$R_{\ind{6r}}^{s\tilde{s}}$ Bragg-reflection polarization matrix for
the cavity in two symmetric configurations.

\subsubsection{$R_{\ind{6m}}^{s\tilde s}$ in the mirror-symmetric cavity}

For the cavity in the mirror-symmetric configuration  we obtain using
Eq.~\eqref{eq054} and substituting the small-$\eta$ expansions of
$a$, $b$, and $d$ from Eq.~\eqref{eq0B040}:
\begin{equation}
\begin{split}
R_{\ind{6m}}^{\sigma\sigma}=&-1-3i\left(2\beta-1\right)\eta+\\
&\left[18\beta(\beta-1) 
  +\alpha^2\left(10+8C_4\right)\right]\eta^2+O(\eta^3),\\[4pt]
R_{\ind{6m}}^{\sigma\pi}=&R_{\ind{6m}}^{\pi\sigma}=2i\alpha S_2\,\eta-12\alpha\beta S_2\,\eta^2+O(\eta^3),\\[4pt]
R_{\ind{6m}}^{\pi\pi}=&-1-3i\left(2\beta+1\right)\eta+\\
&\left[18\beta(\beta+1) 
  +\alpha^2\left(10+8C_4\right)  \right]\eta^2+O(\eta^3).
\label{eq0A1010}
\end{split}
\end{equation}
Here we used the relationship $9C_2^2+S_2^2=5+4C_4$.

The corresponding squared magnitudes are
\begin{equation}
\begin{split}
|R_{\ind{6m}}^{\sigma\sigma}|^2=|R_{\ind{6m}}^{\pi\pi}|^2  =& 1-4\alpha^2S_2^2\eta^2 +O(\eta^3),\\[4pt]
|R_{\ind{6m}}^{\pi\sigma}|^2=& 4\alpha^2S_2^2\eta^2 +O(\eta^3). 
\label{eq0A1015}
\end{split}
\end{equation}
and therefore the squared matrix $\hat{R}_{\ind{6m}}$  is normalized through order $\eta^2$.

Similarly, in the small-deviation limit, 
$\eta \to 0$, the six-crystal reflection matrix $\hat{R}_{\ind{6m}}$ becomes diagonal, since $u \to 0$ and hence the cross-polarization amplitude $|R_{\ind{6m}}^{\pi\sigma}|$ vanishes.

The relative intensity of cross-polarization mixing, quantified by
$\left|R_{\ind{6m}}^{\pi\sigma}\right|^2$ and averaged over the range
$|\eta|\leq 1/4$, is then
\begin{equation}
\langle|R_{\ind{6m}}^{\pi\sigma}|^2\rangle\,\simeq
\frac{\alpha^2S_2^2}{12}
=
\frac{1}{48} \tan^2 2\varphi,
\label{eq0BA1}
\end{equation}
where in the second form we used Eq.~\eqref{eqABphi}.

\subsubsection{$R_{\ind{6r}}^{s\tilde s}$ in the twofold rotationally  symmetric cavity}

For the cavity in the twofold rotationally symmetric configuration  we obtain using
Eq.~\eqref{eq054r} and substituting the small-$\eta$ expansions of
$a$, $b$, and $d$ from Eq.~\eqref{eq0B040}:
\begin{equation}
\begin{split}
R_{\ind{6r}}^{\sigma\sigma}=&-1-3i(2\beta-1)\eta
+\left[
18\beta(\beta-1)+\frac{9}{2}
\right]\eta^2+O(\eta^3),\\[4pt]
R_{\ind{6r}}^{\sigma\pi}=&-R_{\ind{6r}}^{\pi\sigma}
=-3\alpha S_2\,\eta^2+O(\eta^3),\\[4pt]
R_{\ind{6r}}^{\pi\pi}=&-1-3i(2\beta+1)\eta
+\left[
18\beta(\beta+1)+\frac{9}{2}
\right]\eta^2+O(\eta^3).
\end{split}
\label{eq0A1030r}
\end{equation}

The corresponding squared magnitudes are
\begin{equation}
\begin{split}
|R_{\ind{6r}}^{\sigma\sigma}|^2=&\,1+O(\eta^3),\\[4pt]
|R_{\ind{6r}}^{\pi\pi}|^2=&\,1+O(\eta^3),\\[4pt]
|R_{\ind{6r}}^{\pi\sigma}|^2=&|R_{\ind{6r}}^{\sigma\pi}|^2
=9\alpha^2S_2^2 \,\eta^4+O(\eta^5).
                               \label{eq0A1035r}
\end{split}
\end{equation}
Therefore, the squared matrix $\hat{R}_{\ind{6r}}$ is normalized
through order $\eta^2$.

In the small-deviation limit, $\eta\to 0$, the cross-polarization
amplitudes $R_{\ind{6r}}^{\pi\sigma}$ and
$R_{\ind{6r}}^{\sigma\pi}$ vanish as $\eta^2$, and the six-crystal
reflection matrix $\hat{R}_{\ind{6r}}$ becomes diagonal.

The relative intensity of cross-polarization mixing, quantified by
$\left|R_{\ind{6r}}^{\pi\sigma}\right|^2$ and averaged over the range
$|\eta|\leq 1/4$, is then
\begin{equation}
\langle|R_{\ind{6r}}^{\pi\sigma}|^2\rangle\simeq
\frac{9\alpha^2S_2^2}{1280}
=
\frac{9}{5120}\tan^2 2\varphi ,
\label{eq0BA1r}
\end{equation}
where Eq.~\eqref{eqABphi} was used in the last equality.

\subsection{Polarization factor $P^\pi<0$}

Next, we consider the case $P^\pi<0$, corresponding to Bragg angles
in the range $\pi/4 < \theta\leq\pi/2$.

The small-$\eta$ approximations of $v$ and $u$ are
\begin{equation}
\begin{split}
v=&\alpha\eta+i\alpha\beta\,\eta^2+O(\eta^3),\\
u=&-i+\beta\eta+\frac{i}{2}\left(\alpha^2+\beta^2\right)\eta^2+O(\eta^3),
\label{eq0B060}
\end{split}
\end{equation}
for which
\begin{equation}
  |v|\ll |u|.
\label{eq0B080}  
\end{equation}
Since $P^\pi$ and $R^{\pi\pi}$ change sign at $\theta = \pi/4$, the
co-polar amplitude combinations $u$ and $v$, Eq.~\eqref{eq043}  exchange their dominant and
subdominant roles at this angle. 

Substituting into the definitions of $a$, $b$, and $d$, 
in Eq.~\eqref{eq052},
and retaining terms through order $\eta^2$, we obtain
\begin{equation}
\begin{split}
a=&R_{\ind{3}}^{\sigma\sigma}=iC_6-\left(\alpha T+3\beta C_6\right)\eta -\\
&i\left[3\alpha\beta T+3\alpha^2 C_2+K C_6\right]\eta^2+O(\eta^3),\\
b=&R_{\ind{3}}^{\pi\sigma}=-iS_6+3\beta S_6\,\eta+i\left[K S_6+\alpha^2 S_2\right]\eta^2+O(\eta^3),\\
  d=&R_{\ind{3}}^{\pi\pi}=-iC_6+\left(-\alpha T+3\beta C_6\right)\eta +\\
  &i\left[K C_6-3\alpha\beta T+3\alpha^2 C_2\right]\eta^2+O(\eta^3),
\end{split}
\label{eq0B090}
\end{equation}
where
\[
K=\frac{3}{2}\alpha^2+\frac{9}{2}\beta^2.
\]

The corresponding squared moduli through order $\eta^2$ are
\begin{equation}
\begin{split}
|a|^2=|d|^2&=C_6^2+\alpha^2\left(3S_6+2S_2\right)S_6\eta^2+O(\eta^3),
\\
|b|^2&=S_6^2-\alpha^2\left(3S_6^2+2S_2S_6\right)\eta^2+O(\eta^3). 
\end{split}
\label{eq0B100}
\end{equation}
Since $C_6^2+S_6^2=1$,
the squared matrix $\hat{R}_{\ind{3}}$ is normalized through order
$\eta^2$ in agreement with expected unitary behavior of the
three-crystal amplitudes in the $|\eta|\leq 1$ region.

In the next step, these results are used to calculate the small-$\eta$
approximations through order $\eta^2$, for the elements of the
six-crystal  $R_{\ind{6m}}^{s\tilde{s}}$ and
$R_{\ind{6r}}^{s\tilde{s}}$ Bragg-reflection polarization matrix for
the cavity in two symmetric configurations.

\subsubsection{$R_{\ind{6m}}^{s\tilde s}$ in the mirror-symmetric cavity}

For the cavity in the mirror-symmetric configuration  we obtain using
Eq.~\eqref{eq054} and the small-$\eta$ expansions of
$a$, $b$, and $d$ from Eq.~\eqref{eq0B090}:
\begin{equation}
\begin{split}
R_{\ind{6m}}^{\sigma\sigma}=&-1-i\left[2\alpha TC_6+6\beta \right]\eta+\\
&\left[2\alpha^2T^2+12\alpha\beta TC_6+18\beta^2\right]\eta^2+O(\eta^3),\\[4pt]
R_{\ind{6m}}^{\pi\sigma}
=&R_{\ind{6m}}^{\sigma\pi}=2i\alpha TS_6\,\eta-12\alpha\beta TS_6\,\eta^2+O(\eta^3),\\[4pt]
R_{\ind{6m}}^{\pi\pi}
=&-1+i\left[2\alpha TC_6-6\beta \right]\eta+\\
&\left[2\alpha^2T^2-12\alpha\beta TC_6+18\beta^2\right]\eta^2+O(\eta^3),\\[-2pt]
\label{eq0A1020}
\end{split}
\end{equation}
where $T^2=6C_2C_6+2S_6S_2+3$.

The corresponding squared magnitudes are
\begin{equation}
\begin{split}
|R_{\ind{6m}}^{\sigma\sigma}|^2=|R_{\ind{6m}}^{\pi\pi}|^2=& 1-4T^2\alpha^2S_6^2\eta^2+O(\eta^3),\\[4pt]
|R_{\ind{6m}}^{\pi\sigma}|^2 =& 4\alpha^2T^2S_6^2\eta^2 +O(\eta^3). 
\end{split}
\label{eq0A1025}
\end{equation}
Thus the squared matrix $\hat{R}_{\ind{6m}}$ is normalized through order $\eta^2$ in this
case too.

In this case, the relative intensity of the polarization mixing
$|R_{\ind{6m}}^{\pi\sigma}|$ averaged over  $|\eta| \leq 1/4$
range is then 
\begin{equation}
  \langle|R_{\ind{6m}}^{\pi\sigma}|^2\rangle\, \simeq \alpha^2T^2S_6^2/12
  \label{eq0BA2}
\end{equation}

\subsubsection{$R_{\ind{6r}}^{s\tilde s}$ in the twofold rotationally  symmetric cavity}

For the cavity in the twofold rotationally symmetric configuration  we obtain using
Eq.~\eqref{eq054r} and the small-$\eta$ expansions of
$a$, $b$, and $d$ from Eq.~\eqref{eq0B090}:
\begin{equation}
\begin{split}
R_{\ind{6r}}^{\sigma\sigma}=&-C_{12}-2i\left(3\beta C_{12}+\alpha T C_6\right)\eta
+O(\eta^2),\\[4pt]
R_{\ind{6r}}^{\sigma\pi}=&-R_{\ind{6r}}^{\pi\sigma}
=-S_{12}-6i\beta S_{12}\,\eta +O(\eta^2),\\[4pt]
R_{\ind{6r}}^{\pi\pi}=&-C_{12}-2i\left(3\beta C_{12}-\alpha T C_6\right)\eta
+O(\eta^2).
\end{split}
\label{eq0A1040}
\end{equation}
In contrast to the $P^\pi>0$ case, the off-diagonal terms do not
vanish in the limit $\eta\to 0$. Instead,
\begin{equation}
\hat R_{\ind{6r}}(\eta=0)= \hat R_{\ind{6r}}^{(0)} =
\begin{pmatrix}
-\cos 12\varphi & -\sin 12\varphi\\
\sin 12\varphi & -\cos 12\varphi
\end{pmatrix},
\label{eq0A1042}
\end{equation}
so that the residual polarization mixing is determined by
$\sin 12\varphi$.
The corresponding squared magnitudes, to leading order in $\eta$, are
\begin{equation}
\begin{split}
|R_{\ind{6r}}^{\sigma\sigma}|^2=&\,\cos^2 12\varphi+O(\eta),\\[4pt]
|R_{\ind{6r}}^{\pi\pi}|^2=&\,\cos^2 12\varphi+O(\eta),\\[4pt]
|R_{\ind{6r}}^{\pi\sigma}|^2=&|R_{\ind{6r}}^{\sigma\pi}|^2
=\sin^2 12\varphi+O(\eta).
\label{eq0A1045}
\end{split}
\end{equation}
Accordingly, the relative intensity of the polarization mixing
$\left|R_{\ind{6r}}^{\pi\sigma}\right|^2$ averaged over the range
$|\eta|\leq 1/4$ is, to leading order,
\begin{equation}
\langle|R_{\ind{6r}}^{\pi\sigma}|^2\rangle\simeq
\sin^2 12\varphi .
\label{eq0BA2r}
\end{equation}

\section{Cavity eigenpolarizations}
\label{eigen}

Here we analyze the cavity eigenpolarizations and eigenphases in the
small-$\eta$ approximation. The mirror-symmetric cavity is
diagonalized, through order $\eta^2$, by an $\eta$-independent
polarization basis. For the twofold rotationally symmetric cavity,
the intrinsic basis provides the leading eigenbasis for $P^\pi>0$,
whereas circular states diagonalize the zeroth-order matrix for
$P^\pi<0$.

\subsection{Mirror-symmetric   configuration}

Using Eqs.~\eqref{eq0A1010} and \eqref{eq0A1020}, we decompose the
six-crystal reflection  matrix $\hat R_{\ind{6m}}$, through order
$\eta^2$, as
\begin{equation}
\hat R_{\ind{6m}}
=
\begin{pmatrix}
R_{\ind{6m}}^{\sigma\sigma} & R_{\ind{6m}}^{\sigma\pi} \\
R_{\ind{6m}}^{\pi\sigma} & R_{\ind{6m}}^{\pi\pi}
\end{pmatrix}
=R_0 \hat{I}+\lambda \hat{M} + O(\eta^3),
\label{eq0A1030}
\end{equation}
where $\hat{I}$ is the unit matrix. The other components are
discussed  separately in the following for cases $P^\pi>0$ and $P^\pi<0$.

\subsubsection{ $P^\pi>0$}

If $P^\pi>0$,
the average 
$R_0=(R_{\ind{6m}}^{\sigma\sigma}+R_{\ind{6m}}^{\pi\pi})/2$
of the diagonal elements in Eq.~\eqref{eq0A1030} is
\begin{equation}
R_0=-1-6i\beta\eta+\left[18\beta^2+2\alpha^2(9C_2^2+S_2^2) \right]\eta^2,
\end{equation}
\begin{equation}
\lambda = 2i\alpha\eta\left(1+6i\beta\eta\right),
\end{equation}
and
\begin{equation}
\hat{M}=
\begin{pmatrix}
-3C_2 & S_2 \\
S_2 & 3C_2
\end{pmatrix}.
\end{equation}
Thus, to this order, all $\eta$ dependence of the polarization-dependent
part is contained in the scalar factor $\lambda$, while the matrix
$\hat{M}$ is independent of $\eta$.

The eigenvalues of $\hat{M}$ are
\begin{equation}
\mu_\pm=\pm q,
\qquad
q=\sqrt{9C_2^2+S_2^2}. 
\end{equation}
Therefore, the eigenvectors of $\hat R_{\ind{6m}}$ are independent of
$\eta$ through order $\eta^2$ and coincide with the eigenvectors of
$\hat{M}$.

For $S_2\neq 0$, a convenient orthonormal choice is
\begin{equation}
\epp= 
\begin{pmatrix}
\sin\gamma \\
\cos\gamma
\end{pmatrix}, \qquad
\epm= 
\begin{pmatrix}
-\cos\gamma\\
\sin\gamma
\end{pmatrix}.
\label{eq090}
\end{equation}
where
\begin{equation}
  \sin\gamma= \frac{S_2}{\sqrt{2q(q+3C_2)}}, \hspace{2mm}
  \cos\gamma= \frac{q+3C_2}{\sqrt{2q(q+3C_2)}}
\label{eq091}
\end{equation}

These vectors satisfy
\begin{equation}
M\vc{e}_\pm=\mu_\pm\,\vc{e}_\pm,
\qquad
\vc{e}_\pm^\dagger\vc{e}_\pm=1,
\qquad
\epp^\dagger\epm=0 .
\label{eq0910}
\end{equation}
Equivalently, in the intrinsic $\{\esi,\epi\}$ polarization basis  defined in Eq.~\eqref{eq051},
\begin{equation}
\begin{pmatrix}
    \epp \\
    \epm 
\end{pmatrix}\!\!\!
=\hat{W}(\gamma)
\begin{pmatrix}
    \esi \\
    \epi
\end{pmatrix}
,
\hspace{2mm}
\hat{W}(\gamma)
=  
\begin{pmatrix}
\sin\gamma & \cos\gamma \\
-\cos\gamma & \sin\gamma 
\end{pmatrix}
.
\label{eq0915}
\end{equation}    
The inverse matrix $\hat{W}^{-1}(\gamma)=-\hat{W}(-\gamma)$.

The corresponding eigenvalues of the six-crystal reflection matrix are
\begin{equation}
\rho_\pm = R_0\pm \lambda q+O(\eta^3).
\label{eq0940}
\end{equation}
Substituting the expressions for $R_0$, $\lambda$, introducing
\begin{equation}
A_\pm=-6\beta\pm 2\alpha q
\label{eq0942}
\end{equation}
and performing straightforward simplifications we write
\begin{equation}
\rho_\pm=-1+iA_\pm\eta+\frac{A_\pm^2}{2}\eta^2+O(\eta^3).
\label{eq0944}
\end{equation}
To second order in $\eta$,
\begin{equation}
|\rho_\pm|^2=1+O(\eta^3),
\end{equation}
Therefore, both eigenvalues have unit modulus through order $\eta^2$
as it should be for unitary matrices.

We now write the same eigenvalue in the phase form
\begin{equation}
  \rho_\pm(\eta)  =  \exp\left[i\delta_\pm(\eta)\right],
\label{eq0A2030}
\end{equation}
where, to the order considered,
\begin{equation}
  \delta_\pm(\eta)  = \pi-A_\pm\eta+O(\eta^3).
\label{eq0A2040}
\end{equation}
There is no term proportional to $\eta^2$ in the eigenphase.

Thus, within the second-order approximation in $\eta$, the six-crystal
reflection matrix is diagonalized by an $\eta$-independent polarization
basis. Higher-order terms, not included here, could in general introduce
additional polarization-dependent contributions and thereby make the
exact eigenvectors depend on $\eta$.

\subsubsection{$P^\pi<0$}
If $P^\pi<0$,
the average 
$R_0$
of the diagonal elements in Eq.~\eqref{eq0A1030} is\
\begin{equation}
R_0
=
-1-6i\beta\eta+
\left(2\alpha^2T^2+18\beta^2\right)\eta^2,
\end{equation}
\begin{equation}
\lambda
=
2i\alpha T\eta\left(1+6i\beta\eta\right),
\end{equation}
and
\begin{equation}
\hat{M}
=
\begin{pmatrix}
-C_6 & S_6 \\
S_6 & C_6
\end{pmatrix}.
\end{equation}
Thus, also in this case to this order, the entire
polarization-dependent part of $\hat R_{\ind{6m}}$ is an
$\eta$-dependent scalar factor multiplying the fixed matrix $\hat{M}$.

The matrix $\hat{M}$ satisfies $\hat{M}^2=\hat{I}$,
because $C_6^2+S_6^2=1$.
Therefore, the eigenvalues of $\hat{M}$ are
\begin{equation}
\mu_\pm=\pm 1 .
\end{equation}
A convenient orthonormal set of eigenvectors is given by
Eqs.~\eqref{eq090} and \eqref{eq0910}--\eqref{eq0915}, as in the
$P^\pi > 0$ case, but with the angle $\gamma$ given by
\begin{equation}
  \gamma = 3\varphi.
  \label{eq093}
\end{equation}
instead of Eq.~\eqref{eq091}.
Thus, within the second-order approximation in $\eta$, the eigenvectors
of $\hat R_{\ind{6m}}$ are also in this case independent of $\eta$.

The corresponding eigenvalues of the reflection matrix are given by
Eqs.~\eqref{eq0940} and \eqref{eq0944} as in the $P^\pi>0$ case,
but with $A_\pm$ given by
\begin{equation}
  A_\pm=-6\beta\pm 2\alpha T
\label{eq0A2050}  
\end{equation}
instead of Eq.~\eqref{eq0942}.

Because, both eigenvalues have unit modulus through order $\eta^2$,
the eigenvalues can likewise be expressed via
eigenphases $\delta_\pm(\eta)$ as in
Eqs.~\eqref{eq0A2030}-\eqref{eq0A2040}, but in this case 
$A_\pm$ is given by 
Eq.~\eqref{eq0A2050}. 

This result shows that also for $P^\pi<0$, to second order in
the deviation parameter, the six-crystal reflection matrix is
diagonalized by the $\eta$-independent polarization eigenvectors
$\vc{e}_\pm$. Higher-order terms, not included here, could in general
introduce additional polarization- dependent structures and make the
exact eigenvectors depend on $\eta$.

\subsubsection{Physical meaning}

The fact that the eigenvectors are independent of $\eta$, whereas the
eigenvalues themselves are $\eta$ dependent but have unit modulus, has
a clear physical meaning. In the considered approximation, the
six-reflection cavity behaves as a lossless retarder with fixed
eigenpolarizations but $\eta$-dependent eigenphases.

For the incident field \eqref{eq0510}  with  arbitrary  polarization,
\[
   \vc{E}_{\indrm{i}}
 =  E_{\indrm{i}}^{\sigma}\esi +
  E_{\indrm{i}}^{\pi}\epi =  
  c_+\epp + c_-\epm 
\]
\[
  \begin{split}
    c_+=& ~E_{\indrm{i}}^{\sigma}\sin\gamma  +E_{\indrm{i}}^{\pi} \cos\gamma  \\
       c_-=& -E_{\indrm{i}}^{\sigma}\cos\gamma +
             E_{\indrm{i}}^{\pi} \sin\gamma ,  
  \end{split}
  \]
the field after the round trip in the cavity is
\[
  \vc{E}_{\rm out}  =  c_+\rho_+(\eta)\epp +  c_-\rho_-(\eta)\epm .
\]
Because $|\rho_\pm|=1$, the magnitudes of the modal coefficients
$c_+$ and $c_-$ are unchanged. However, the relative phase
\[
  \Delta\delta(\eta)=  \delta_+(\eta)-\delta_-(\eta)
\]
depends on $\eta$. Therefore, unless the incident field is one of the
eigenpolarizations, the output polarization state generally changes with
$\eta$ through this changing relative phase.

\subsection{Twofold rotationally symmetric configuration}

We now consider the six-crystal reflection matrix
$\hat R_{\ind{6r}}$ for the cavity in the twofold rotationally
symmetric configuration. Using the small-$\eta$ expansions derived in
Sec.~\ref{smalleta} for the matrix elements of
$\hat R_{\ind{6r}}$, we examine whether the corresponding cavity
eigenpolarizations are also independent of $\eta$ through second
order.

\subsubsection{$P^\pi>0$}

For $P^\pi>0$, using Eq.~\eqref{eq0A1030r} for the matrix elements of
$\hat R_{\ind{6r}}$, we write
\begin{equation}
\hat R_{\ind{6r}}
=
\begin{pmatrix}
R_{\ind{6r}}^{\sigma\sigma} & R_{\ind{6r}}^{\sigma\pi} \\
R_{\ind{6r}}^{\pi\sigma} & R_{\ind{6r}}^{\pi\pi}
\end{pmatrix}
=R_0 \hat{I}+\hat{\Delta} + O(\eta^3),
\label{eq0A2080}
\end{equation}
where
\begin{equation}
R_0
=
-1-6i\beta\eta+
\left(
18\beta^2+\frac{9}{2}
\right)\eta^2.
\end{equation}
and
\begin{equation}
\hat{\Delta}
=
\begin{pmatrix}
3i\,\eta-18\beta \,\eta^2 & -3\alpha S_2\,\eta^2 \\
3\alpha S_2\,\eta^2 & -3i\,\eta+18\beta\,\eta^2
\end{pmatrix}.
\end{equation}

Because the off-diagonal elements of $\hat R_{\ind{6r}}$ are of order
$\eta^2$, whereas the diagonal splitting already appears in order
$\eta$, the leading polarization-dependent part is diagonal in the
intrinsic basis $\{\esi,\epi\}$ for $C_2\neq 0$. Therefore, within the
present approximation, the cavity eigenvectors may be identified with
the intrinsic basis vectors,
\begin{equation}
\epp=\esi=
\begin{pmatrix}
1\\
0
\end{pmatrix},
\qquad
\epm=\epi=
\begin{pmatrix}
0\\
1
\end{pmatrix},
\label{eq0A2090}
\end{equation}
up to corrections of higher order in $\eta$.

The corresponding eigenvalues are
\begin{equation}
\rho_+ = R_{\ind{6r}}^{\sigma\sigma}+O(\eta^3),
\qquad
\rho_- = R_{\ind{6r}}^{\pi\pi}+O(\eta^3),
\end{equation}
that is,
\begin{equation}
\begin{split}
\rho_+=&-1-3i(2\beta-1)\eta+\\
&\left[18\beta^2-18\beta -\alpha^2S_2^2\right]\eta^2+O(\eta^3),\\[4pt]
\rho_-=&-1-3i(2\beta+1)\eta+\\
&\left[18\beta^2+18\beta-\alpha^2S_2^2\right]\eta^2+O(\eta^3).
\end{split}
\label{eq0A2100}
\end{equation}

Introducing
\begin{equation}
A_\pm=-6\beta\pm 3,
\label{eq0A2110}
\end{equation}
the eigenvalues may be written, as in the mirror-symmetric case, in the
phase form
\begin{equation}
\rho_\pm(\eta)=\exp\left[i\delta_\pm(\eta)\right],
\end{equation}
with
\begin{equation}
\delta_\pm(\eta)=\pi-A_\pm\eta+O(\eta^3).
\end{equation}

Thus, for $P^\pi>0$, the six-crystal reflection matrix in the twofold
rotationally symmetric configuration is diagonalized, within the
second-order approximation in $\eta$, by the intrinsic polarization
basis itself, which in this case coincides with the cavity
eigenpolarization basis $\{\epp,\epm\}$. The cavity therefore behaves
as a lossless retarder with fixed eigenpolarizations and
$\eta$-dependent eigenphases.

\subsubsection{$P^\pi<0$}

For $P^\pi<0$, the situation is qualitatively different. Using
Eq.~\eqref{eq0A1040}, the matrix $\hat R_{\ind{6r}}$ has nonvanishing
off-diagonal terms already at zeroth order in $\eta$. It is therefore
convenient to separate the zeroth-order part from the $\eta$-dependent
corrections:
\begin{equation}
\hat R_{\ind{6r}}
=
\hat R_{\ind{6r}}^{(0)}
+\eta\,\hat R_{\ind{6r}}^{(1)}
+O(\eta^2),
\label{eq0A2140}
\end{equation}
where $\hat R_{\ind{6r}}^{(0)}$ defined in Eq.~\eqref{eq0A1042} may be written as
\begin{equation}
\hat R_{\ind{6r}}^{(0)}
=
-\cos12\varphi\,\hat I
-\sin12\varphi
\begin{pmatrix}
0&1\\
-1&0
\end{pmatrix},
\end{equation}

In the practically interesting cases $\sin 12\varphi \not = 0$, which
exclude extreme cavity configurations $\theta=\pi/4$ and
$\theta=\pi/2$, its eigenvalues are
\begin{equation}
\rho_\pm^{(0)} 
=-\exp(\mp i12\varphi),
\label{eq0A2160}
\end{equation}
with corresponding eigenvectors
\begin{equation}
\epp=
\frac{1}{\sqrt{2}}
\begin{pmatrix}
1\\
-i
\end{pmatrix},
\qquad
\epm=
\frac{1}{\sqrt{2}}
\begin{pmatrix}
1\\
i
\end{pmatrix}.
\label{eq0A2170}
\end{equation}
Thus, the eigenpolarizations are circular in the
intrinsic basis, are independent of $\varphi$ or $\theta$, and
equivalently can be defined in the intrinsic $\{\esi,\epi\}$ polarization basis 
\begin{equation}
\begin{pmatrix}
    \epp \\
    \epm 
\end{pmatrix}
=\hat{W}
\begin{pmatrix}
    \esi \\
    \epi
\end{pmatrix}
,
\quad
\hat{W}
= \frac{1}{\sqrt 2} 
\begin{pmatrix}
1 & -i \\
1 & i
\end{pmatrix}.
\label{eq0915r}
\end{equation}    
with
\begin{equation}
\hat{W}^{-1}
= \frac{1}{\sqrt 2} 
\begin{pmatrix}
1 & 1 \\
i & -i
\end{pmatrix}.
\label{eq09155r}
\end{equation}    
However, in contrast to the mirror-symmetric case,  the first-order correction from Eq.~\eqref{eq0A1040} 
\begin{equation}
\hat R_{\ind{6r}}^{(1)}
=
6i\beta\hat R_{\ind{6r}}^{(0)}
-2i\alpha T C_6
\begin{pmatrix}
1&0\\
0&-1
\end{pmatrix}.
\label{eq09157r}
\end{equation}
is not proportional to the same matrix structure as $\hat R_{\ind{6r}}^{(0)}$.
In the zeroth-order circular basis, the diagonal matrix elements of
the traceless first-order correction vanish. This correction therefore
couples the two circular modes but does not shift their eigenvalues in
first order for nondegenerate values of $\sin12\varphi$. The
first-order eigenvalue correction is consequently determined by the
term proportional to $\hat R_{\ind{6r}}^{(0)}$ and is common to both
modes:
\begin{equation}
\rho_\pm(\eta)=-\exp(\mp i12\varphi)\left[1+6i\beta\eta\right] +O(\eta^2).
\label{eq0A2190}
\end{equation}
Thus the corresponding first-order eigenphases are
\begin{equation}
\delta_\pm(\eta) = \pi\mp12\varphi+6\beta\eta+O(\eta^2).
\end{equation}
The $\alpha T$ term in Eq.~\eqref{eq09157r} primarily changes the eigenvectors at first order rather than producing a first-order eigenphase splitting in the generic nondegenerate case.

For $P^\pi<0$, the circular basis $\{\epp,\epm\}$ diagonalizes the
zeroth-order reflection matrix for generic values
$\sin12\varphi\neq0$. The full cavity eigenpolarizations generally
acquire $\eta$-dependent corrections, because the traceless part of
the first-order correction couples the two circular modes. At the
special points $\sin12\varphi=0$, the zeroth-order matrix is
degenerate and the eigenpolarizations must be determined from the
first-order correction.

\end{document}